\documentclass[a4paper]{jpconf}
\usepackage[hiresbb]{graphicx}
\usepackage[english]{babel}
\usepackage{amsmath,amsthm}
\usepackage{multirow}
\usepackage{bbold}
\usepackage{tikz}
\usepgflibrary{arrows.meta}

\newcommand{\nn}{\nonumber}

\newcommand{\dd}{\mathrm{d}} 

\newcommand{\M}[1]{\mathbf{#1}}

\newcommand{\RR}{\mathbb{R}}
\newcommand{\CC}{\mathbb{C}}

\newcommand{\ee}{\mathrm{e}}
\newcommand{\ii}{\mathrm i}

\DeclareMathOperator{\sign}{\mathrm{sign}}

\DeclareMathOperator{\IM}{\mathrm{Im}}

\newcommand{\PHI}{\phi}

\newtheorem{theorem}{Theorem}
\newtheorem{defn}[theorem]{Definition}

\begin{document}

\title{Pseudo-hermitian random matrix theory: a review}
\author{Joshua Feinberg$^{1,2,3}$ and Roman Riser$^{1,2}$}

\address{$^1$ Department of Mathematics and}
\address{$^2$ Haifa Research Center for Theoretical Physics and Astrophysics, University of Haifa, Haifa 31905, Israel}
\address{$^3$ https://orcid.org/0000-0002-2869-0010}

\begin{abstract}
We review our recent results on pseudo-hermitian random matrix theory which were hitherto presented in various conferences and talks. (Detailed accounts of our work will appear soon in separate publications.) Following an introduction of this new type of random matrices, we focus on two specific models of matrices which are pseudo-hermitian with respect to a given indefinite metric $\M{B}$. Eigenvalues of pseudo-hermitian matrices are either real, or come in complex-conjugate pairs. The diagrammatic method is applied to deriving explicit analytical expressions for the density of eigenvalues in the complex plane and on the real axis, in the large-$N$, planar limit. In one of the models we discuss, the metric $\M{B}$ depends on a certain real parameter $t$. As $t$ varies, the model exhibits various 'phase transitions' associated with eigenvalues flowing from the complex plane onto the real axis, causing disjoint eigenvalue support intervals to merge. Our analytical results agree well with presented numerical simulations.
\end{abstract}

\pagestyle{plain}

\section{Introduction}
PT-symmetric quantum mechanics (PTQM)  \cite{BB} and its broader applications (see \cite{CMB} for recent reviews) have been at the focus of intensive and prolific research activity during the past quarter of century. Broadly speaking, the Hilbert space of a PT-symmetric quantum mechanical model is endowed with a non-trivial metric operator, with respect to which the hamiltonian is hermitian. Hamiltonians of PTQM models with proper metrics are sometimes referred to as {\it quasi-hermitian} \cite{Dieudonne, stellenbosch}. They are diagonalizable, and their spectrum is essentially real, because they are similar to a conventionally hermitian hamiltonian. In contrast, hamiltonians of PTQM models with {\it indefinite} metrics are referred to as {\it pseudo-hermitian} \cite{Froissart, Mostafazadeh}, and their eigenvalues are either real or come in  complex-conjugate pairs. For a mathematically precise summary of the nomenclature of quasi-hermiticity and pseudo-hermiticity see  \cite{Fring-Assis}.

Quasi-hermitian matrices can be thought of as truncated quasi-hermitian linear operators. In \cite{FRDecember, MK} we found it useful to further distinguish {\it strictly}-quasi-hermitian matrices within the broader class of quasi-hermitian ones: Strictly-quasi-hermitian matrices are hermitian with respect to positive definite (and therefore invertible) metrics. In contrast, merely quasi-hermitian matrices are associated with non-negative non-invertible metrics.  

An interesting strictly-quasi-hermitian random matrix model was introduced in \cite{JK}. These authors fixed a metric, and took the hamiltonian as random, with the aim of studying numerically the dependence of the average density of eigenvalues and level spacing statistics on the metric. Yet another interesting example of a strictly-quasi-hermitian random matrix model, akin to the Dicke model of superradiance, was provided by \cite{DGK}, in which a numerical study of the level spacing distribution was carried out. 

Upon truncation to finite vector spaces, pseudo-hermitian operators turn into pseudo-hermitian matrices. See \cite{KA} for a recent discussion of (real asymmetric) pseudo-hermitian random matrices.

In this paper we introduce a family of pseudo-hermitian random matrix models and review our analytical and numerical results for the distribution of real and complex eigenvalues. A detailed account of these investigations will appear soon separately \cite{FRB,FR}. \\

Let us start with the definitions of strict quasi-hermiticity and pseudo-hermiticity: 
\begin{defn}[strict quasi-hermiticity]\label{def:quasi}	
Let $\M{B}$ and $\PHI$ be $N\times N$ matrices where $\PHI$ is a complex matrix and $\M{B}$ a hermitian, positive definite \emph{metric}. We call $\PHI$ strictly-quasi-hermitian with respect to the metric $\M{B}$ if it fulfills the \emph{intertwining relation}
\begin{equation}\label{eq:inter}
\PHI^\dagger \M{B}=\M{B} \PHI. 
\end{equation}
\end{defn}
The definition means that $\PHI$ is hermitian with respect to the metric $\M{B}$. For $\M{B}=\mathbb{1}$ it reduces to ordinary hermiticity.

Given the metric $\M{B}$, the general solution of the intertwining relation \eqref{eq:inter} is 
\begin{equation} \label{eq:A}
\PHI=\M{A}\M{B},
\end{equation}
where $A=A^\dagger$ is an arbitrary hermitian matrix \cite{JK}.  Thus, given $\M{B}$, there are $N^2$ independent quasi-hermitian matrices with respect to $\M{B}$.

An important corollary is that for a (positive definite) metric, the spectrum of $\PHI$ is always real. This can be seen easily when we write 
\begin{equation}
\PHI=\sqrt{\M{B}}^{\ -1}\underbrace{\sqrt{\M{B}}\M{A}\sqrt{\M{B}}}_{\text{hermitian}} \sqrt{\M{B}}
\end{equation}
where it is obvious that $\PHI$ is similar to a hermitan matrix, provided  $\sqrt\M{B}$ is hermitian, which requires that $\M{B}$ be positive. Spectra of strictly-quasi-hermitian matrices 
were studied in \cite{FRDecember, MK, JK,DGK}. \\

When we relax the condition that $\M{B}$ has to be positive definite, we come to the definition of pseudo-hermiticity:
\begin{defn}[pseudo-hermiticity]\label{def:pseudo}
Let $\M{B}$ and $\PHI$ be $N\times N$ matrices where $\PHI$ is a complex matrix and $\M{B}$ an \emph{indefinite} (hermitian) metric. We call $\PHI$ pseudo-hermitian with respect to the indefinite metric $\M{B}$ if it fulfills the \emph{intertwining relation}
\begin{equation}\label{eq:inter2}
\PHI^\dagger \M{B}=\M{B} \PHI. 
\end{equation}
\end{defn}

The general solution of the intertwining relation \eqref{eq:inter2} is still given by \eqref{eq:A} where $\M{A}$ is an arbitrary hermitian matrix \cite{FRB}. However, contrary to the strictly-quasi-hermitian case, the spectrum of pseudo-hermitian matrices need no longer be purely real, and may contain pairs of complex-conjugated eigenvalues.

As a motivation, let us first give a simple physical example where strictly-quasi-hermitian matrices appear \cite{FRDecember, MK}. Consider a {\it highly-connected} mechanical system with $N$ degrees of freedom with generalized coordinates $q$, executing small oscillations about a stable equilibrium state $q_0$. Its equation of motion is 
\begin{equation}\label{eq:motion}
\M{M} \ddot  x= -\M{K} x,
\end{equation}
where $\M{M}$ is the hermitian and strictly positive mass matrix, $\M{K}$ the hermitian and positive definite matrix of
spring constants and $x=q-q_0$ denotes the deviation from the equilibrium point $q_0$. The Lagrangian of the system is given by
\begin{equation}
L=\dot q^{T} \M{a}(q) \dot q - U(q),
\end{equation} 
where $\M{a}(q)$ is the strictly positive definite metric of the configuration space, with $\M{M}=\M{a}(q_0)$, $U(q)$ is the potential energy, and 
\begin{equation}
K_{ij}=\left.\frac{\partial^2 U}{\partial q_i \partial q_j}\right|_{{q_0}},
\end{equation}
the Hessian at the minimum.

The solution of the equation of motion \eqref{eq:motion} can be found in terms of harmonic eigenmodes
\begin{equation}
x=A e^{i\omega t},
\end{equation}
where $A$ is the amplitude vector and $\omega$ the corresponding oscillation frequency. The eigenmode equation reads
\begin{equation}\label{eq:eigenmode}
(-\omega^2 \M{M}+\M{K})A=0.
\end{equation}
All frequencies must be real, of course, since we are dealing with small oscillations around a stable minimum. 

Alternatively, we can write the Eq. \ref{eq:eigenmode} as
\begin{equation}
\PHI A=\omega^2 A,
\end{equation}
where $\PHI=\M{M}^{-1}\M{K}$ and $\PHI^\dagger=\M{K}\M{M}^{-1}$. Since in general, $\M{M}$ and $\M{K}$ do not commute, $\PHI$ is clearly not hermitian, but it is strictly-quasi-hermitian in respect to the metric $\M{M}$ as it fulfills the intertwining relation  $\PHI^\dagger \M{M}=\M{M}\PHI$. Therefore it is an example where strictly-quasi-hermitian matrices appear.

For very large mechanical systems with high connectivity (all particles are coupled to each other) this problem naturally lends itself to analysis in terms of random matrices. The matrices $M$ and $K$ have to be drawn from sensible uncorrelated
probability distributions of positive definite matrices.  

In \cite{FRDecember, MK}, $\M{K}$ was drawn from the Wishart ensemble with variance $\sigma^2$ and $\M{M}$ from the shifted Wishart ensemble with variance $\sigma'^2$ with a shift $m_0$, i.e.
\begin{align}
\M{M}= \M{C}^\dagger \M{C} + m_0,\qquad\qquad \M{K}= \tilde{\M{C}}^\dagger \tilde{\M{C}}
\end{align}
where $\M{C}$, $\tilde{\M{C}}$ are complex $N\times N$ matrices drawn from the probability distribution functions (PDF)
\begin{align}
P_M(\M{C})=Z^{-1} \ee^{-\frac{N}{\sigma^2} \tr \M{C}^\dagger \M{C} }, \qquad P_K(\tilde{\M{C}})=Z^{-1} \ee^{-\frac{N}{\sigma'^2} \tr \tilde{\M{C}}^\dagger \tilde{\M{C}} }.
\end{align}
Using free probability theory, one can obtain the eigenvalue statistics of $\PHI=\M{M}^{-1}\M{K}$. \cite{FRDecember, MK} \\

In this review, we want to discuss what happens when the metric is no longer positive, i.e.~we will concentrate on the case of indefinite metric $\M{B}$, and therefore pseudo-hermitian matrix $\PHI$. Such a scenario may arise as a modification of the mechanical model discussed above, in which the system undergoes a structural transition, rendering the mass matrix $\M{M}$ ``tachyonic", that is, possess negative eigenvalues. Alternatively, some springs may lose their elasticity, making  the equilibrium point $q_0$ unstable (a saddle point of the potential $U(q)$), which in turn renders the matrix $\M{K}$ non-positive definite. Yet another important motivation for indefinite metrics originates from quantum mechanical systems with broken $PT$ symmetry \cite{BB, CMB}. The hamiltonian $H$ of a  $PT$-symmetric system satisfies the intertwining relation $H^\dagger {\cal PC} = {\cal PC} H$, where the metric is the product of the parity operator ${\cal P}$ and the so-called  ${\cal C}$ operator. As explained in Chapter 3 of \cite{CMB}, the operator ${\cal C}$ is hermitian, commutes with both $H$ and ${\cal PT}$, and like parity, squares to the identity. The strictly quasi-hermitian case (positive definite ${\cal PC}$) corresponds to unbroken $PT$ symmetry. Here each eigenstate of $H$ is also an eigenstate of the 
${\cal PT}$ operator and the corresponding eigenenergy is real. The pseudo-hermitian case (indefinite ${\cal PC}$), on the other hand, corresponds to broken  $PT$ symmetry. In this case, eigenstates of $H$ are either eigenstates of ${\cal PT}$ as well, with real eigenenergies, or come in doublets with complex-conjugate eigenenergies, and the member eigenstates of each doublet are mapped onto each other by the ${\cal PT}$ operator. (In the broken phase, there should be at least one such pair of complex conjugate eigenvalues.) We refer the interested reader to Sections 1.1 and 1.2 in \cite{CMB} for an elementary lucid explanation of $PT$ symmetry breaking in qualitative physical terms. Now, imagine  a highly structured complicated $PT$-symmetric system. Such a system, like the complex mechanical system discussed above, naturally lends itself to analysis in terms of random matrix theory. 

The structure of the rest of this paper is as follows: In Section \ref{sec:defn} we will define our pseudo-hermitian random matrix model regarding a given metric $\M{B}$. In Section \ref{sec:methods} we will briefly review the methods we have used to find an explicit analytic result of the eigenvalues in the limit when $N\rightarrow\infty$. More details will appear in \cite{FRB}. In Section \ref{sec:modelB} we will discuss the results when the metric is of the form $\M{B}=\mathrm{diag}(1,\ldots,1,-1,\ldots,-1)$. The main results of this section are the density of eigenvalues on the real axis given by \eqref{eq:rhoreal} and the parametrization (see Eq. \eqref{eq:rpm}) of the complex domain where the density is uniform. The analytical formulas are confronted with various results from numerical simulations. Finally, in Section \ref{sec:modelt} we will generalize the metric to the form $\M{B}=\mathrm{diag}(1,\ldots,1,t,\ldots,t)$ where we concentrate on the case when $t<0$. This model will show various phase transition in the density of real eigenvalues as well as in the domain of complex eigenvalues. We will summarize the results for the criticality of the parameters in a phase diagram. A brief discussion section concludes this review. 

\section{Indefinite Metric: Model of Pseudo-Hermitian Random Matrices}\label{sec:indefiniteMetric}
\subsection{Definition of the Model}\label{sec:defn}
We will study a random matrix ensemble which is pseudo-hermitian (see Definition \ref{def:pseudo}) with respect to (w.r.t.) a deterministic, indefinite metric $\M{B}$. If $\PHI$ is an element of the set of matrices which are pseudo-hermitian w.r.t.~the metric $\M{B}$, it fulfills the intertwining relation \eqref{eq:inter2}. As we have mentioned in the introduction, the general solution is given by $\PHI=\M{A}\M{B}$ where $\M{A}=\M{A}^\dagger$ is an arbitrary hermitian matrix (see Eq. \eqref{eq:A}). 

In order to define an ensemble of pseudo-hermitian random matrices w.r.t.~the given metric $\M{B}$, we are free to choose a probability distribution on $\M{A}$ in \eqref{eq:A}. Since $\M{A}$ has to be hermitian, a natural choice is to draw $\M{A}$ from the Gaussian Unitary Ensemble (GUE), i.e.~we put on $\M{A}$ the PDF
\begin{equation}\label{eq:GUE}
\tilde{P}(\M{A})=\frac{1}{\tilde{Z}_N} \ee^{-\frac{N m^2}{2} \tr \M{A}^2},
\end{equation}
where $m>0$ is a parameter and $\tilde{Z}_N$ is the normalization constant. The PDF \eqref{eq:GUE} induces the following PDF on $\PHI$,
\begin{equation}
P(\PHI)=\frac{1}{Z_N} \ee^{-\frac{N m^2}{2} \tr (\M{B}^{-2} \PHI^\dagger \PHI) } \delta(\phi^\dagger \M{B}-\phi \M{B}),
\end{equation}
where $Z_N$ is another normalization constant.

We will see that $\PHI$ can contain complex and real eigenvalues. The main goal of this paper is to analyze the density of eigenvalues of $\PHI$ on the real axis and to find the domain of complex eigenvalues in the limit $N\rightarrow\infty$.

\subsection{Methods for analyzing the large $N$ limit}\label{sec:methods}
The averaged density of eigenvalues of the pseudo-hermitian matrix $\PHI$ we can calculate from the the resolvent of $\PHI=\M{A}\M{B}$ (Green's function)
\begin{equation}\label{eq:Green}
G(w)=\left\langle \frac{1}{N} \tr \frac{1}{w-\M{A}\M{B}} \right\rangle,
\end{equation}
where the brackets denote averaging of $\M{A}$ over the GUE ensemble. We will see that in the large $N$ limit we can obtain $G(w)$ explicitly in closed form.

Averaging over $\M{A}$ becomes simpler if one can avoid the product of matrices $\M{A}\M{B}$ by the method introduced in \cite{BJN}, where the authors define a $2N \times 2N$ block matrix $\M{H}$,
\begin{equation}
\M{H}=\left(\begin{array}{cc} 
0&\M{A}\\\M{B}&0 
\end{array}\right),
\end{equation}
with its resolvent 
\begin{equation}
\frac{1}{z-\M{H}}=\left(\begin{array}{cc} 
\frac{z}{z^2-\M{A}\M{B}}& \frac{1}{z^2-\M{A}\M{B}}\M{A}\\
\frac{1}{z^2-\M{B}\M{A}}\M{B}& \frac{z}{z^2-\M{B}\M{A}}
\end{array}\right),
\end{equation}
and its Green's function
\begin{equation}\label{eq:Gtilde}
\widetilde{G}(z)= \left\langle \frac{1}{2N} \tr \frac{1}{z-\M{H}} \right\rangle.
\end{equation}
From the upper left block of the resolvent of $\M{H}$ we can get the desired Green's function \eqref{eq:Green} when we divide by z and substitute $w=z^2$. Helpful will be the fact that the block of the matrix $\M{H}$, which includes the
random matrix $\M{A}$, is decoupled from the block with the deterministic metric $\M{B}$.

Since $\M{H}$ is typically non-hermitian, it might have complex eigenvalues. In the large $N$ limit, they can be dense in a two-dimensional domain in $\CC$. Therefore $\widetilde{G}$ is not an analytic function of $w$: It is not just a function of $w$ but also of $w^*$.\footnote{This can be seen for example in the complex Ginibre ensemble, where the eigenvalue density in the large $N$ limit is uniformly supported on the unit disk. Its Green's function is given by
\begin{equation*}
G_{\textrm{Gin}}(z)=\left\{ \begin{array}{ll} z^* &\text{if } |z|<1,\\ \frac{1}{z} &\text{if } |z|\ge 1. \end{array}\right. 
\end{equation*}
The density of eigenvalues can be obtained from $G_{\textrm{Gin}}(z)$ by the anti-holomorphic derivative, 
\begin{equation*}
\varrho_{\textrm{Gin}}(z)=\frac{1}{\pi} \frac{\partial}{\partial z^*} G_{\textrm{Gin}}(z).
\end{equation*}}

We will overcome the difficulty mentioned above by the method of \emph{Hermitization} (see \cite{FZ,JNGZ,CW,E97}). For this we have to double the size of the matrix again, so now we have to deal with the
resolvent of a $4N \times 4N$ block matrix
\begin{equation}\label{eq:G} 
\hat{\mathcal{G}}=\left[\left(\begin{array}{cc}
\eta&z\\z^*&\eta 
\end{array}\right)-\left(\begin{array}{cc} 
0&\M{H}\\\M{H}^\dagger&0 
\end{array}\right)\right]^{-1}=\left(\begin{array}{cccc}
\eta&0&z&-\M{A}\\
0&\eta&-\M{B}&z\\
z^*&-\M{B}&\eta&0\\
-\M{A}&z^*&0&\eta
\end{array}\right)^{-1}.
\end{equation}
Notice that all the elements of the matrix on the right-hand side of Eq.~\eqref{eq:G} are $N\times N$ blocks. $\hat{\mathcal{G}}$ is the resolvent of the $4N\times 4N$ matrix 
\begin{equation}\label{eq:herm}
\left(\begin{array}{cc} 
0&z-\M{H}\\z^*-\M{H}^\dagger&0 
\end{array}\right).
\end{equation}
Important is the fact that the matrix given by \eqref{eq:herm} is \emph{hermitian}. Therefore, off the real axis in the complex $\eta$-plane,  $\hat{\mathcal{G}}$ is a holomorphic function of the complex spectral parameter $\eta$. This will allow us to make a series expansion in $1/\eta$, which we will do in a diagrammatic way.

Following 't Hooft, we will use Feynman diagrams with double lines \cite{thooft} in the planar limit\cite{QFTNut}, when $N\rightarrow\infty$. We will calculate the planar limit by the following expansion in Feynman diagrams. We expand the resolvent $\hat{\mathcal{G}}$ in powers of bare propagators
\begin{equation}
\hat{\mathcal{G}}_0=\hat{\mathcal{G}}_{|_{A=0}}=\begin{tikzpicture}[scale=0.3,line width=0.4pt,baseline=-3pt]
\path (0,0) circle [radius=1cm];
\draw (-2,0) -- (2,0);
\path[tips, -{Latex[scale length=1.5,scale width=1.125]}] (-2,0) -- (0,0);
\end{tikzpicture}
\end{equation}
and rearrange the perturbative expansion in terms of the self-energy
\begin{equation}
\hat{\Sigma}\quad=\quad\begin{tikzpicture}[scale=0.2,line width=0.4pt,baseline=0.25cm]
\path (0,0) circle [radius=1cm] node[above={0.15cm}] {$\hat{\Sigma}$};
\draw (-5,0) arc [start angle=180, end angle=0, radius=5cm] -- cycle;
\end{tikzpicture}\quad
\end{equation}
in the following way:
\begin{align*}
\begin{tikzpicture}[scale=0.45,line width=0.4pt,baseline=-3pt]
\draw (0,0) circle [radius=1cm];
\draw (-2,0) -- (-1,0) (1,0) -- (2,0);
\path (0,0) node {$\langle\hat{\mathcal{G}}\rangle$};
\path[tips, -{Latex[scale length=1.5,scale width=1.125]}] (-2,0) -- (-1.25,0);
\path[tips, -{Latex[scale length=1.5,scale width=1.125,reversed]}] (2,0) -- (1.25,0);
\end{tikzpicture}
\quad&=\quad
\begin{tikzpicture}[scale=0.45,line width=0.4pt,baseline=-3pt]
\path (0,0) circle [radius=1cm];
\draw (-2,0) -- (2,0);
\path[tips, -{Latex[scale length=1.5,scale width=1.125]}] (-2,0) -- (0,0);
\end{tikzpicture}
\quad+\quad
\begin{tikzpicture}[scale=0.45,line width=0.4pt,baseline=0.15cm]
\draw (-1.5,0) arc [start angle=180, end angle=0, radius=1.5cm] -- cycle;
\draw (-2.5,0) -- (-1.5,0) (1.5,0) -- (2.5,0);
\path[tips, -{Latex[scale length=1.5,scale width=1.125]}] (-2.5,0) -- (-1.75,0);
\path[tips, -{Latex[scale length=1.5,scale width=1.125,reversed]}] (2.5,0) -- (1.75,0);
\path (0,0) node[above={-0.05cm}] {$\hat{\Sigma}$};
\path (0,0) circle [radius=1cm];
\end{tikzpicture}&\!\!\!\!\!+\quad
\begin{tikzpicture}[scale=0.45,line width=0.4pt,baseline=0.15cm]
\draw (-1.5,0) arc [start angle=180, end angle=0, radius=1.5cm] -- cycle;
\draw (3,0) arc [start angle=180, end angle=0, radius=1.5cm] -- cycle;
\draw (-2.5,0) -- (-1.5,0) (1.5,0) -- (4,0) (6,0) -- (7,0) ;
\path[tips, -{Latex[scale length=1.5,scale width=1.125]}] (-2.5,0) -- (-1.75,0);
\path[tips, -{Latex[scale length=1.5,scale width=1.125]}] (1.5,0) -- (2.75,0);
\path[tips, -{Latex[scale length=1.5,scale width=1.125,reversed]}] (7,0) -- (6.25,0);
\path (0,0) node[above={-0.05cm}] {$\hat{\Sigma}$};
\path (4.5,0) node[above={-0.05cm}] {$\hat{\Sigma}$};
\path (0,0) circle [radius=1cm];
\end{tikzpicture}
\quad+\quad\ldots
\end{align*}
This rearrangement is allowed because $\hat{\mathcal{G}}$ is an analytic function of $\eta$, and therefore this Born series is convergent. This would not have been possible before the procedure of Hermitization.

$\hat{\Sigma}$ is the sum over all 1-quark irreducible diagrams\cite{QFTNut}.
We can express $\hat{\Sigma}$ in terms of the connected cumulants of the distribution of $\M{A}$ and the full propagator $\langle\hat{\mathcal{G}}\rangle$. Since $\M{A}$ is from the GUE ensemble, there is only one connected cumulant.

In the planar limit we arrive at the following expression for
the self energy, the so called \emph{gap equation},
\begin{equation}\label{eq:gap}
\begin{tikzpicture}[scale=0.3,line width=0.4pt,baseline=0.75cm]
\path (0,0) circle [radius=1cm] node[above={0.4cm}] {$\hat{\Sigma}$};
\draw (-5,0) arc [start angle=180, end angle=0, radius=5cm] -- cycle;
\end{tikzpicture}
\quad=\quad
\centering\begin{tikzpicture}[scale=0.35,line width=0.4pt,baseline=0.75cm]
\draw (0,0) circle [radius=1cm];
\draw (5.5,0) arc [start angle=0, end angle=180, radius=5.5cm];
\draw (-1,0) -- (-5,0) arc [start angle=180, end angle=0, radius=5cm] -- ((1,0);
\filldraw[fill=white] (0,5.2) circle [radius=1cm]; 
\path (0,0) node {$\langle\hat{\mathcal{G}}\rangle$};
\path (0,5.2) node {\tiny $\langle \! A^2 \! \rangle_{\!\mathrm{c}}$};
\path[tips, -{Latex[scale length=2,scale width=1.5]}] (-5,0) -- (-2.8,0);
\path[tips, -{Latex[scale length=2,scale width=1.5,reversed]}] (5,0) -- (2.8,0);
\path[tips, -{Latex[scale length=2,scale width=1.5]}](-5.5,0) arc [start angle=180, end angle=135, radius=5.5cm];
\path[tips, -{Latex[scale length=2,scale width=1.5,reversed]}](-5,0) arc [start angle=180, end angle=135, radius=5cm];
\path[tips, -{Latex[scale length=2,scale width=1.5,reversed]}](5.5,0) arc [start angle=0, end angle=45, radius=5.5cm];
\path[tips, -{Latex[scale length=2,scale width=1.5]}](5,0) arc [start angle=0, end angle=45, radius=5cm];
\end{tikzpicture}
\quad=\quad\left(\begin{array}{cccc} 
\overline{44}&0&0&\overline{41}\\
0&0&0&0\\
0&0&0&0\\
\overline{14}&0&0&\overline{11}
\end{array}\right),
\end{equation}
where $\overline{\alpha \beta}=N^{-1} \langle \tr \hat{\mathcal{G}}_{\alpha\beta}\rangle$ are block traces with $\alpha,\beta=1,2,3,4$. Notice that the block traces are numbers, therefore the right-hand side of Eq.~\eqref{eq:gap} is a $4\times 4$ matrix. The appearance of many zeros in this matrix is due to the decoupling of the random blocks from the deterministic
blocks, which we have mentioned in the end of the paragraph below \eqref{eq:Gtilde}. 

We are left to find out the four quantities $\overline{44}$, $\overline{41}$, $\overline{14}$ and $\overline{11}$. After taking $\eta\rightarrow 0$, the block traces are uniquely determined by a self-consistent equation, the \emph{Schwinger-Dyson equation}, 
\begin{equation}
\langle \hat{\mathcal{G}}\rangle=\frac{1}{\hat{\mathcal{G}}_0^{-1}-\hat{\Sigma}}.
\end{equation}
All other blocks of $\langle\hat{\mathcal{G}}\rangle$ can be found in terms of the four quantities $\overline{11},\overline{14},\overline{41},\overline{44}$. These four quantities are determined from their self-consistent equations (see e.g.~Eq.~\eqref{eq:cubic} for $\overline{41}$). In particular, we find for the $\overline{31}$ block, 
\begin{align}
\left \langle \frac{1}{N} \tr \hat{\mathcal{G}}_{31} \right\rangle&=\left \langle \frac{1}{N} \tr \frac{z}{z^2-\M{A}\M{B}} \right\rangle\nn \\ 
&=\frac{m^2 z}{N} \tr \frac{\M{B}^{-1}(\overline{14}-m^2 w^* \M{B}^{-1})}{\overline{11} \cdot \overline{44}-(\overline{41}-m^2w \M{B}^{-1})(\overline{14}-m^2w^* \M{B}^{-1})},
\end{align}
where we remind the reader that $w=z^2$. 

It admits two different kinds of solutions for the Green's function, a
\emph{holomorphic} one and a \emph{non-holomorphic} solution. From the holomorphic solution we will be able to obtain the eigenvalue density on the real axis, while the non-holomorphic solution will give the complex eigenvalues. 

The block traces $\overline{11}$ and $\overline{44}$ are either both zero, or non-zero. If $\overline{11}=\overline{44}=0$, then $\overline{41}$ is a holomorphic function of $w$ away from the real axis, as well as the Green's function
\begin{equation}\label{eq:Gsol}
G(w)=\left\langle \frac{1}{N} \tr \frac{1}{w-\M{A}\M{B}} \right\rangle=-\frac{m^2}{N} \tr \frac{\M{B}^{-1}}{\overline{41}-m^2 w \M{B}^{-1}}.
\end{equation}
It accounts for the purely real part of the spectrum of $\PHI=\M{A}\M{B}$. 

On the other-hand, if $\overline{11}$ and $\overline{44}$ are non-zero, all quantities are non-holomorphic functions of $w$. This solution accounts for eigenvalues of $\PHI$ in the complex plane. The boundary of the two-dimensional domain occupied by the complex eigenvalues is obtained by setting $\overline{11}=\overline{44}=0$ in the non-holomorphic gap equations.
	
For any positive definite metric $\M{B}$, one can consistently show that only the holomorphic solution exists, while for an indefinite metric, the holomorphic and non-holomorphic solution co-exist. So far we have not specified the deterministic metric $\M{B}$. To proceed further, we will choose a specific form of metric which we will do in the following sections. Then we can find an explicit solution for the density of real eigenvalues and the domain of complex eigenvalues.

\subsection{Density of Eigenvalues for $\M{B}=\mathrm{diag}(1,\ldots,1,-1,\ldots,-1)$}\label{sec:modelB}

In this section we will make a particular choice of the indefinite metric $\M{B}$. For simplicity we will choose $\M{B}$ to be diagonal and of the form
\begin{equation}\label{eq:metricB}
\M{B}=\mathrm{diag}(\underbrace{1,\ldots,1}_{k},\underbrace{-1,\ldots,-1}_{N-k}),
\end{equation}
with $1<k<N$. We will define the fraction of ones on the diagonal of $\M{B}$,
\begin{equation}\label{eq:lambda}
\lambda=\frac{k}{N},
\end{equation}
which will be especially convenient since we usually consider the limit $N\rightarrow \infty$. Notice that from the general solution \eqref{eq:A} we can see that the model has the obvious symmetry 
\begin{equation}\label{eq:symB}
\lambda \mapsto 1-\lambda.
\end{equation}

We will now come back to the holomorphic solution which we have mentioned at the end of Section \eqref{sec:methods}. For the particular choice of metric $\M{B}$ given in \eqref{eq:metricB}, the holomorphic gap equation reduces to the cubic equation
\begin{equation}\label{eq:cubic}
m^2 b + \frac{\lambda}{b+w}+\frac{1-\lambda}{b-w}=0, 
\end{equation}
where
\begin{equation}\label{eq:b}
b=-m^{-2}\overline{41}.
\end{equation}	
Taking the solution of the cubic equation \eqref{eq:cubic} for $b$ with the asymptotic behavior
\begin{equation}
b(w)\sim \frac{1-2\lambda}{m^2 w},\qquad (w\rightarrow \infty),
\end{equation}
and substitute it in Eq.~\eqref{eq:Gsol} using \eqref{eq:b}, we can get the density of real eigenvalues from the discontinuities of the Green's function $G(w)$ across the real axis. We find that in the large $N$ limit the density of real eigenvalues of $\PHI$ is supported on the interval $[-a,a]$ where it is given by
\begin{align}\label{eq:rhoreal}
\rho^{(1)}(x)&=\frac{1}{2\pi} \lim_{\epsilon\searrow 0} \IM \Big[G(x- \ii \epsilon)-G(x + \ii \epsilon)\Big]\nn\\
&=\sign(1-2\lambda) \, \frac{\left|\xi-\sqrt{\Delta} \right|^{2/3} - \left|\xi+\sqrt{\Delta} \right|^{2/3}}{\sqrt{3} \cdot 2^{2/3}\cdot 6\pi m^2 x}, \qquad |x|<a,
\end{align}
where 
\begin{align}
a&= \Bigg( \tfrac{3|1-2\lambda|^{2/3}\,\big[\,\big(1-2\sqrt{\lambda(1- \lambda)} \big)^{1/3}+\big(1+2\sqrt{\lambda(1- \lambda)} \big)^{1/3}\,\big]+2}{2m^2}\Bigg)^{\!{}_{1/2}}, \label{eq:endpoints}\\
\xi&=\xi(x)=-27 m^4 (1-2\lambda)x,\\
\intertext{and the discriminant}
\Delta&=\Delta(x)=\xi^2+4\cdot 27 m^6(1-m^2 x^2)^3.
\end{align}
\begin{figure}
	\begin{center}
		\includegraphics[width=\textwidth]{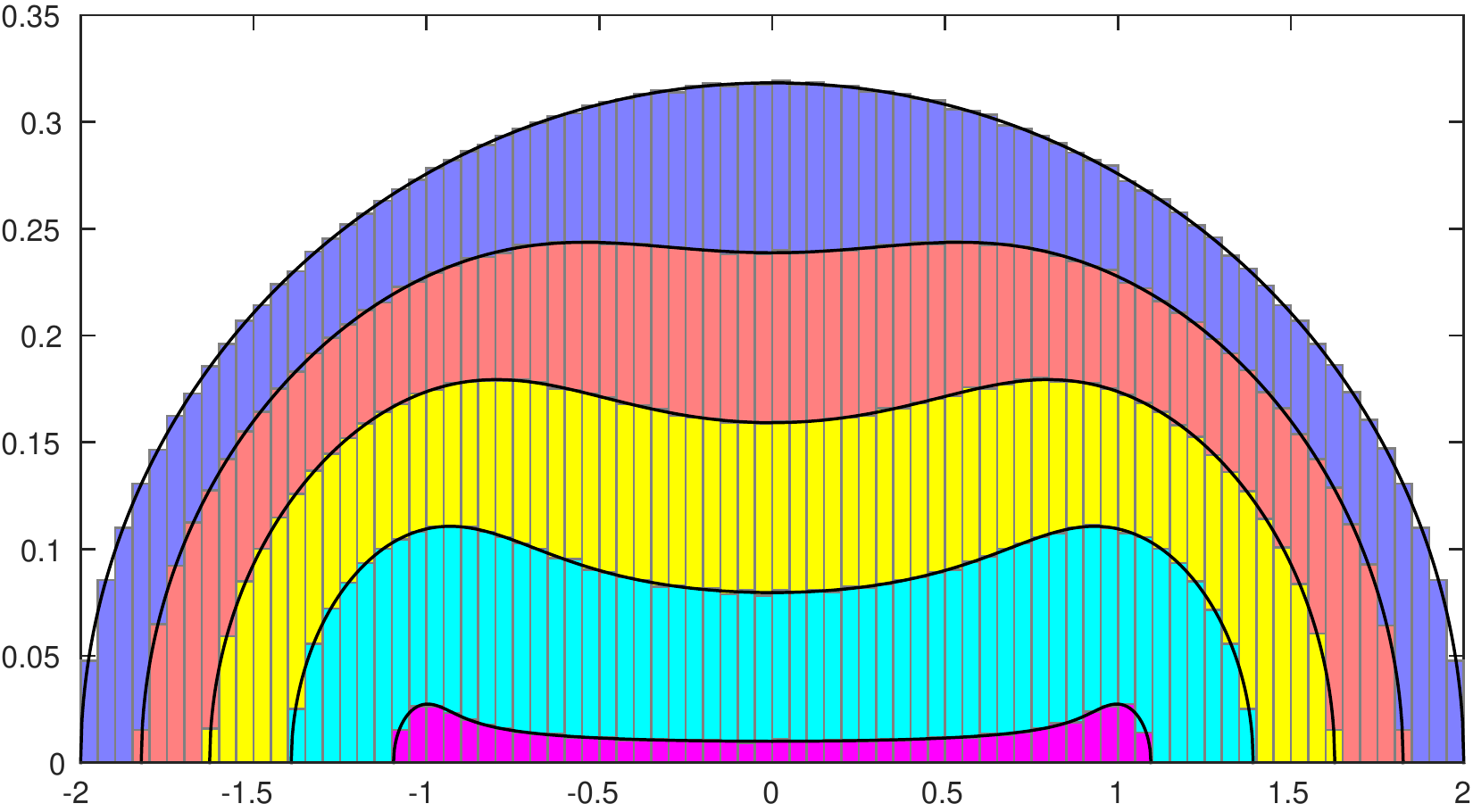} 
	\end{center} 
	\caption{Normalized histograms of real eigenvalues from a single sample using a numerical simulation with $N=32768$. Different color represent different values of $\lambda$: 31/64, 3/8, 1/4, 1/8, 1/32768 (ordered as density increases). The solid black line shows the corresponding large $N$ theoretical prediction $\rho^{(1)}(x)$ given by \eqref{eq:rhoreal}.}\label{fig:rhoreal}
\end{figure}
Note that for $x$ going to zero, the density behaves like
\begin{equation}
\lim_{x\rightarrow 0}\rho^{(1)}(x)=\frac{m|1-2\lambda|}{\pi}.
\end{equation}

In Fig.~\ref{fig:rhoreal} we show normalized histograms obtained from numerically generated samples for various $\lambda$'s. For each such value we have used a single sample of large matrix size with $N=32768$. The histograms fit well with the solid black lines from the corresponding theoretical limits $N\rightarrow\infty$ given by Eq.~\eqref{eq:rhoreal}. That a single sample is suitable to represent the averaged density, comes from the self-averaging effect. This is typical for the eigenvalue density of large random matrices.

The histogram in magenta shows the result for $\lambda=31/64$ which is close to the degenerated case where the theoretical density \eqref{eq:rhoreal} predicts that there is no density of real eigenvalues. When lambda decreases from $1/2$, the density of real eigenvalues increases. The blueish histogram shows the case where $\lambda$ is close to zero. For $\lambda=0$ it is obvious from the model, that $\PHI=\M{A}$ which has been drawn from GUE, so the semi-circle law is no surprise. 

Since we have normalized the density such that all real and all complex eigenvalues together sum up to one, integrating $\rho^{(1)}(x)$ gives the fraction of real eigenvalues as a function of $\lambda$. We find from \eqref{eq:rhoreal} that the fraction of real eigenvalues is
\begin{equation} \label{eq:vshaped}
\int_{-a}^a \rho^{(1)}(x) \dd x= |1-2\lambda|,
\end{equation}
\begin{figure}
\begin{center}
\includegraphics[width=\textwidth]{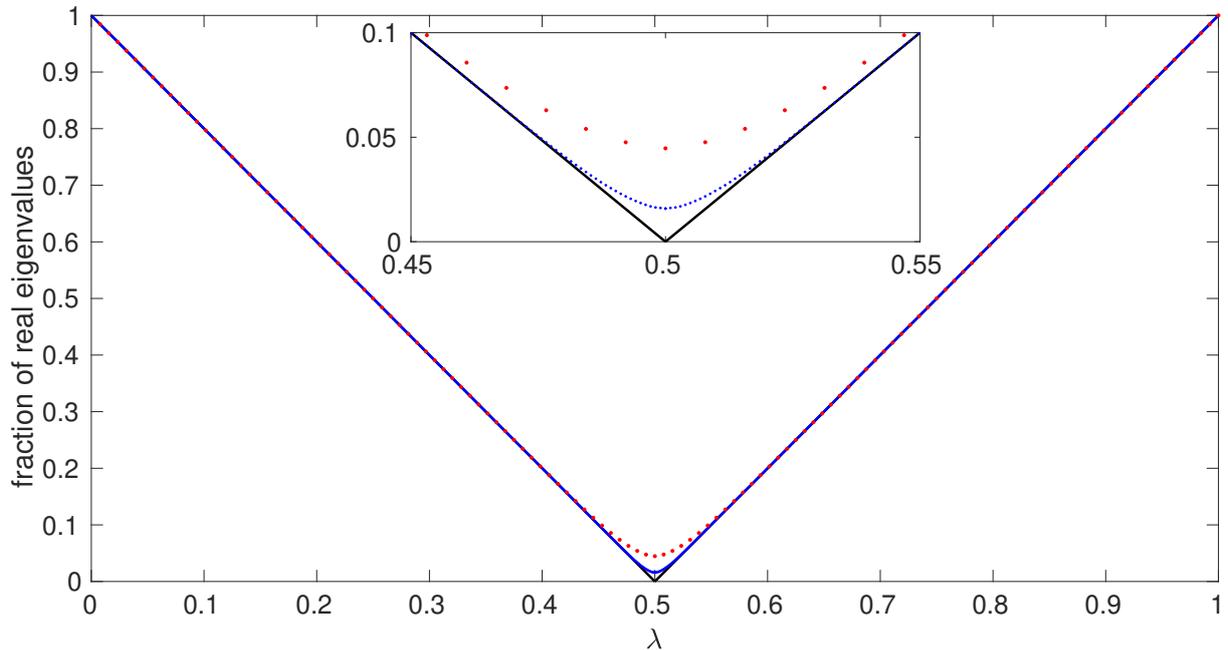}
\end{center} 
\caption{Fraction of real eigenvalues as a function of $\lambda$. Solid black line shows the theoretical large $N$ prediction given by \eqref{eq:vshaped}, while the dots represent numerical simulations with $N=128$ averaged over 500000 samples (red) and $N=1024$ using 2000 samples (blue). Inset: Magnification near $\lambda=1/2$.
}\label{fig:vshaped}
\end{figure}
which is consistent with the symmetry of the model \eqref{eq:symB}. In Fig.~\ref{fig:vshaped} this result has been compared with numerical simulations for matrix sizes $N=128$ and $N=1024$ averaged over many samples. For $\lambda$ away from $1/2$ the convergence is exponentially fast. Near $1/2$, we can see a small deviation for finite $N$. The theoretical large $N$ prediction, actually is a lower bound for the number of real eigenvalues, even for finite $N$ and for each sample. This follows from a special case of a purely algebraic theorem \cite{carlson}.

Next we want to discuss what we find for the choice of metric $\M{B}$ given by \eqref{eq:metricB} for the non-holomorphic solution mentioned at the end of section \ref{sec:methods} which accounts for the complex eigenvalues. As we have discussed before, in this case, both $\overline{11}$ and $\overline{44}$ are non-zero and the Green's function $G(w)$ is non-holomorphic. More precisely, when $N\rightarrow\infty$, $G(w)$ is non-holomorphic in the domain which is densely filled by complex eigenvalues and holomorphic outside. At the boundary of this domain, $G(w)$ changes continuously to the Green's function we have obtained from the holomorphic solution. Therefore we can find the boundary of the complex domain by setting $\overline{11}=\overline{44}=0$ in the non-holomorphic gap equation. In the large $N$ limit, we find for our particular choice of metric the following parametrization in polar coordinates,
\begin{equation}\label{eq:rpm}
r_\pm(\theta)=\frac{1}{\sqrt{2}m}\left( 1\pm \sqrt{1-\left( \frac{\sin \theta_0}{\sin \theta } \right)^2} \right)^{1/2}\!\!\!\!\!\!\!\!, 
\end{equation}
when $\sin^2\theta\ge \sin^2 \theta_0$ with $\sin \theta_0=|2\lambda-1|$. $r_+$ is the part of the boundary farther from the origin, and $r_-$ the closer one.	
\begin{figure}
	\begin{center}
		\includegraphics[width=0.495\textwidth]{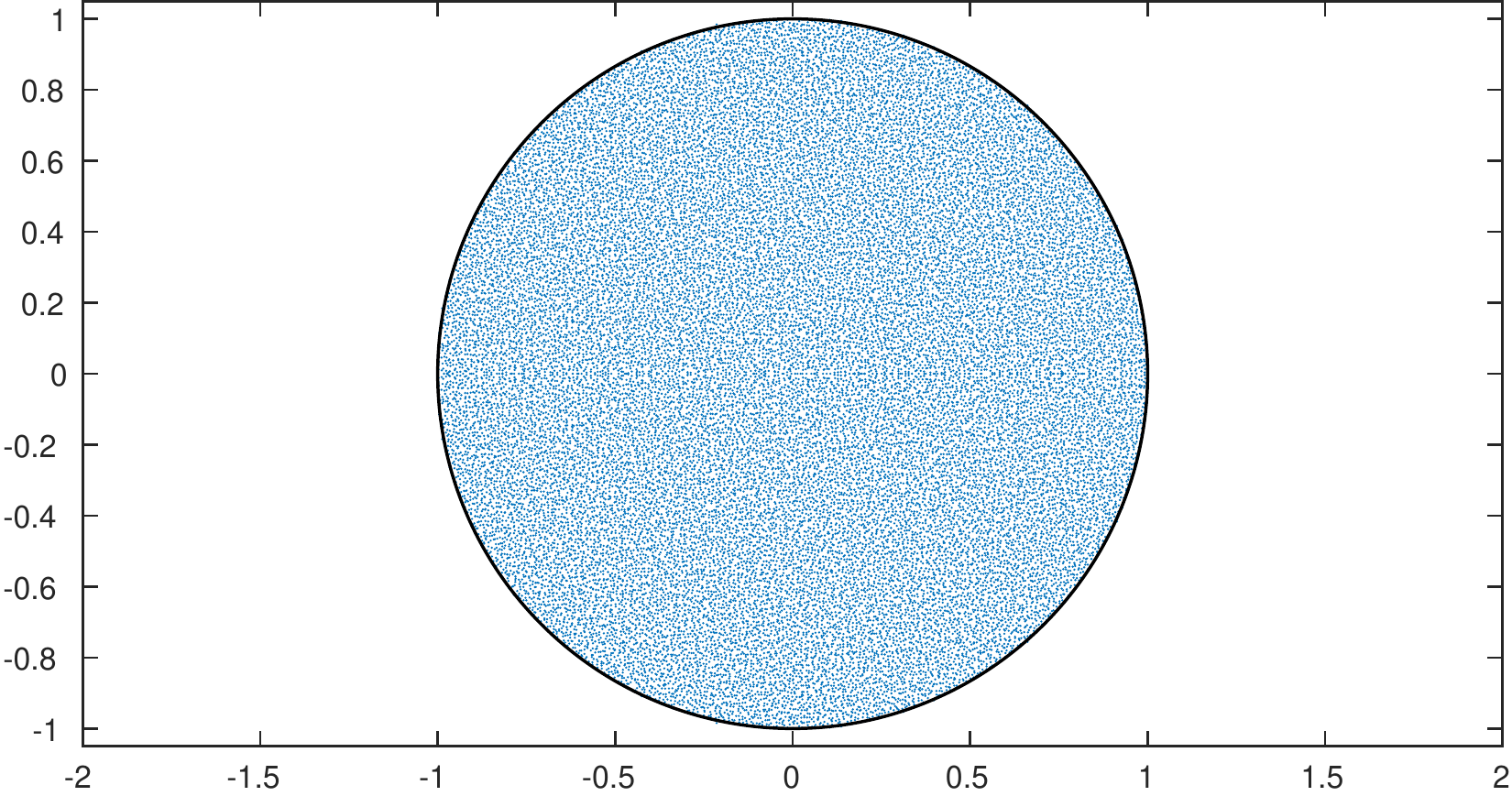}\hfill \includegraphics[width=0.495\textwidth]{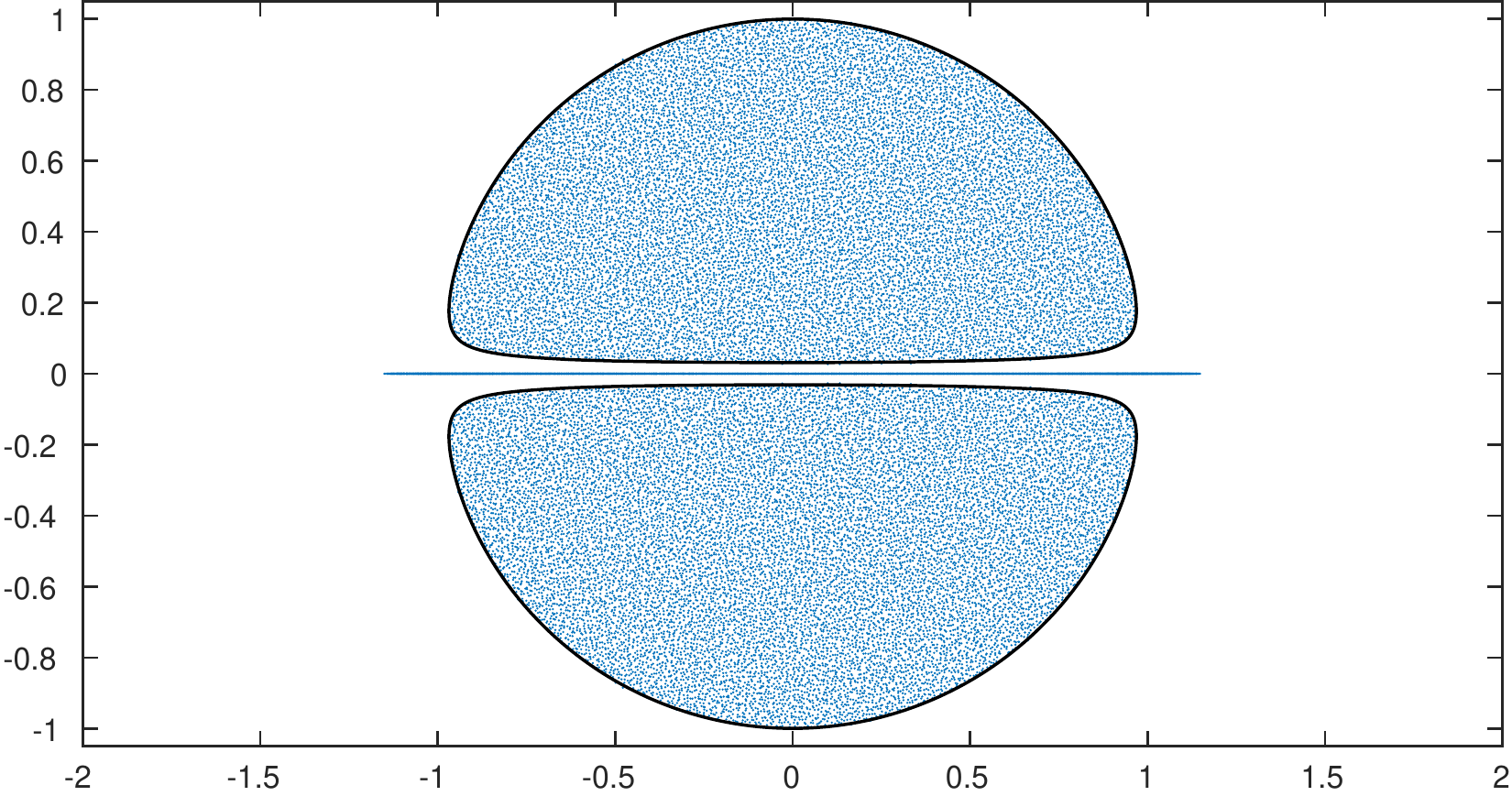}      
		\includegraphics[width=0.495\textwidth]{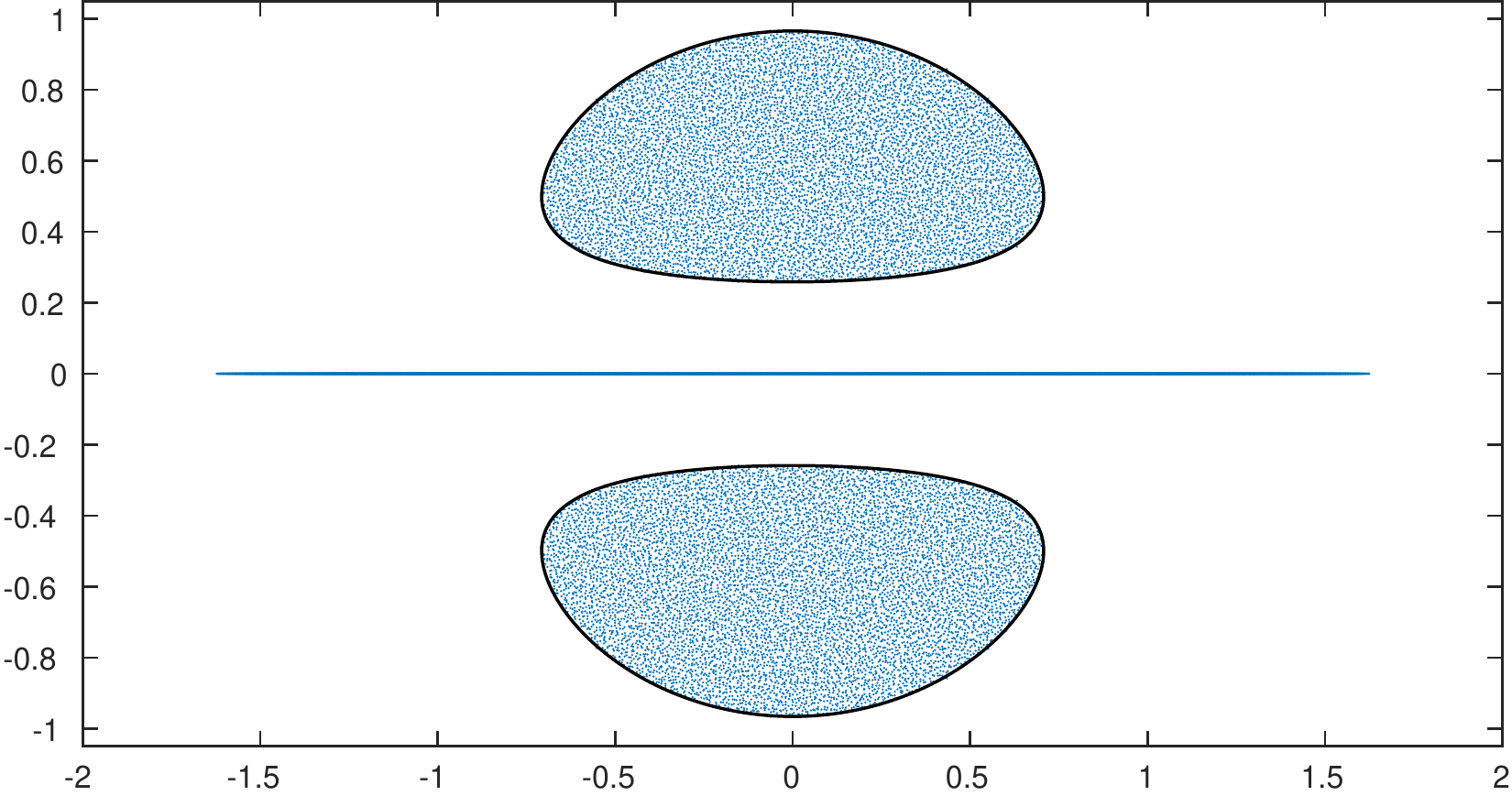}\hfill \includegraphics[width=0.495\textwidth]{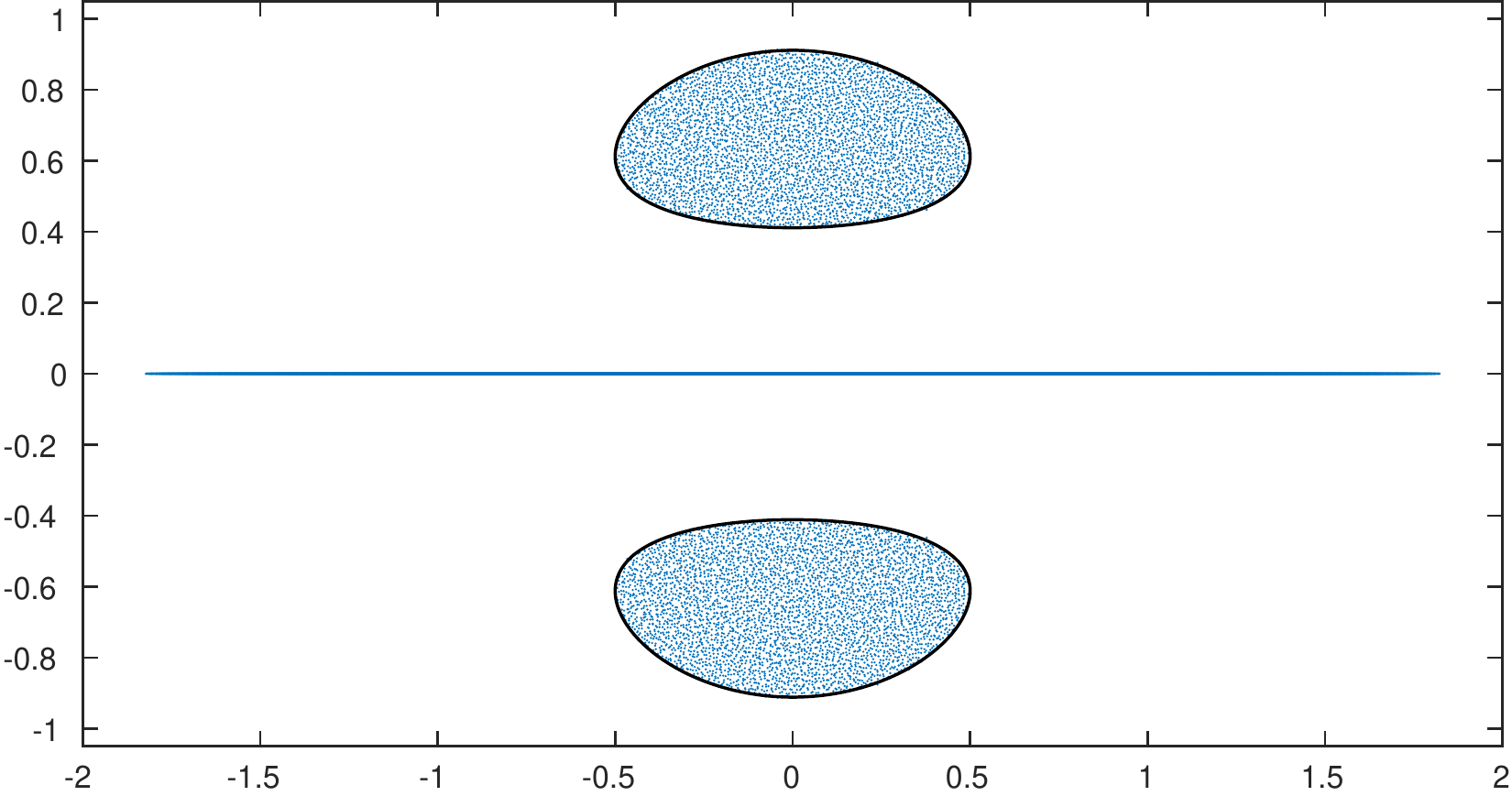}
	\end{center}
	\caption{The blue dots represent the eigenvalues of $\PHI=\M{A}\M{B}$ obtained from numerical simulations using a single realization with $N=32768$ for various values of $\lambda$: 1/2, 15/32, 1/4, 1/8 (from upper left to lower right). The solid black line shows the theoretical boundary  when $N\rightarrow\infty$ given by \eqref{eq:rpm}. }\label{fig:scatterB}
\end{figure}

In Fig.~\ref{fig:scatterB} we illustrate the results for various values of $\lambda$ by scatter plots. The blue dots correspond to the eigenvalues of $\PHI$ obtained from single samples of matrices of size $N=32768$. Since the scatter plots have been produced by a single sample, it demonstrates well how uniformly the eigenvalues are distributed which is due to the repulsive force among them. The solid black lines represent the theoretical boundaries given by Eq.~\eqref{eq:rpm}.

On the upper left plot, we see the degenerated case $\lambda=1/2$ for which the theory predicts null density of real eigenvalues. On the remaining scatter plots we see also eigenvalues on the real axis where the blue dots appear as a line since they are very dense as there are $|N-2k|=N|1-2\lambda|$ real eigenvalues. For fixed $\lambda$, the typical distance between eigenvalues on the real lines is of order $N^{-1}$ while the mean distance between nearest neighbors of complex eigenvalues in the bulk is of order $N^{-1/2}$.   

For each sample, the eigenvalue distribution is symmetric w.r.t.~the real axis. This is due to the fact that the characteristic polynomial of $\det(z-\PHI)$ has real coefficients and the complex eigenvalues appear in conjugated pairs. It is a manifestation of PT symmetry. The eigenvalue distribution has an additional symmetry regarding the imaginary axis, but only after averaging over all samples.

We can calculate the area of the complex domain, the two blobs parameterized by \eqref{eq:rpm}, by the following integral
\begin{equation}\label{eq:area}
2 \int_{r_-}^{r_+} \int_{\theta_0-\pi}^{2\pi-\theta_0} r \dd r \dd \theta=  \int_{\theta_0-\pi}^{2\pi-\theta_0} \left(r_+^2(\theta)-r_-^2(\theta)\right)\dd \theta=\frac{(1-|1-2\lambda|)\pi}{m^2}.
\end{equation}
\begin{figure}
\begin{center}
\includegraphics[width=0.495\textwidth]{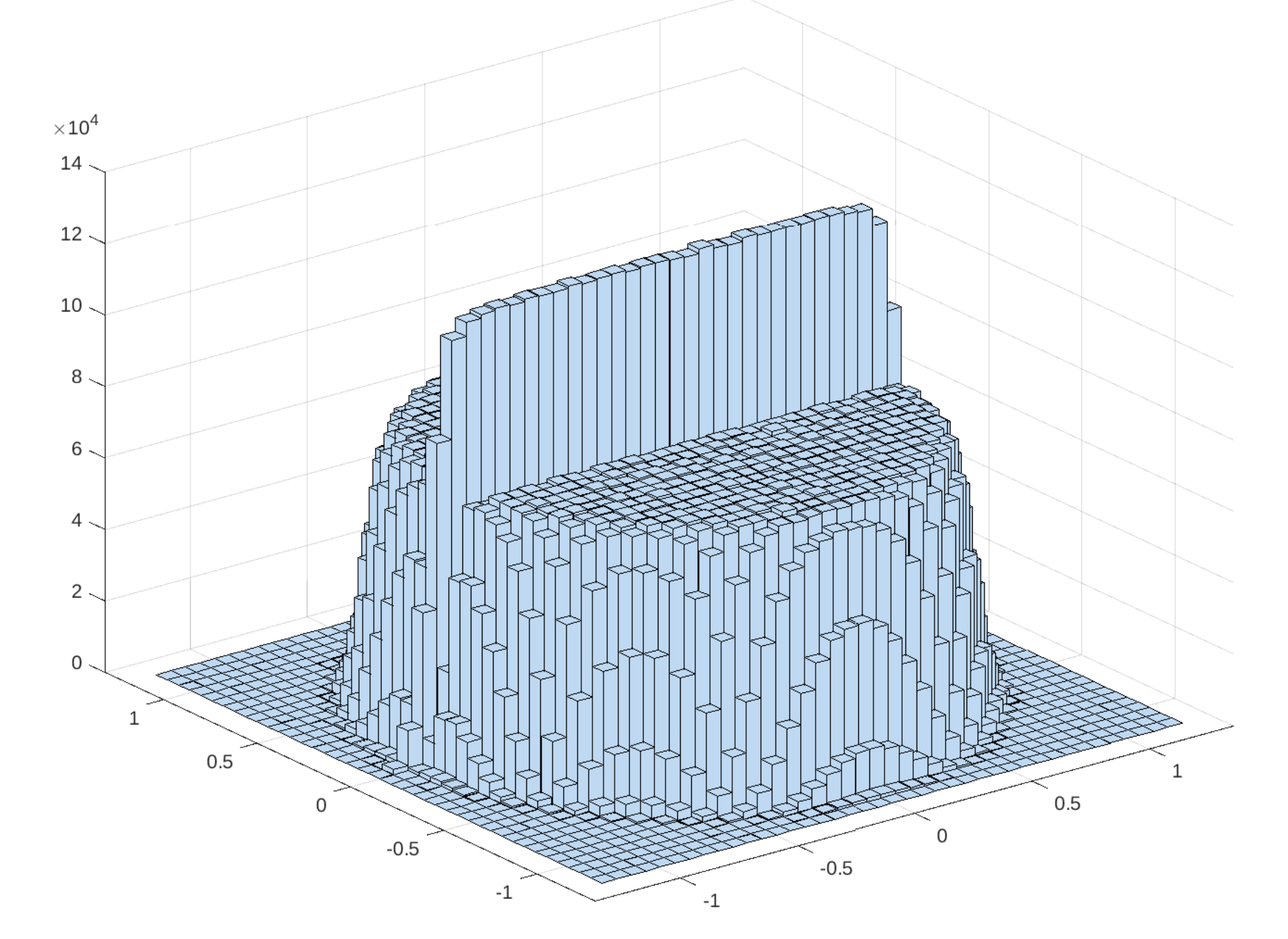}\hfill \includegraphics[width=0.495\textwidth]{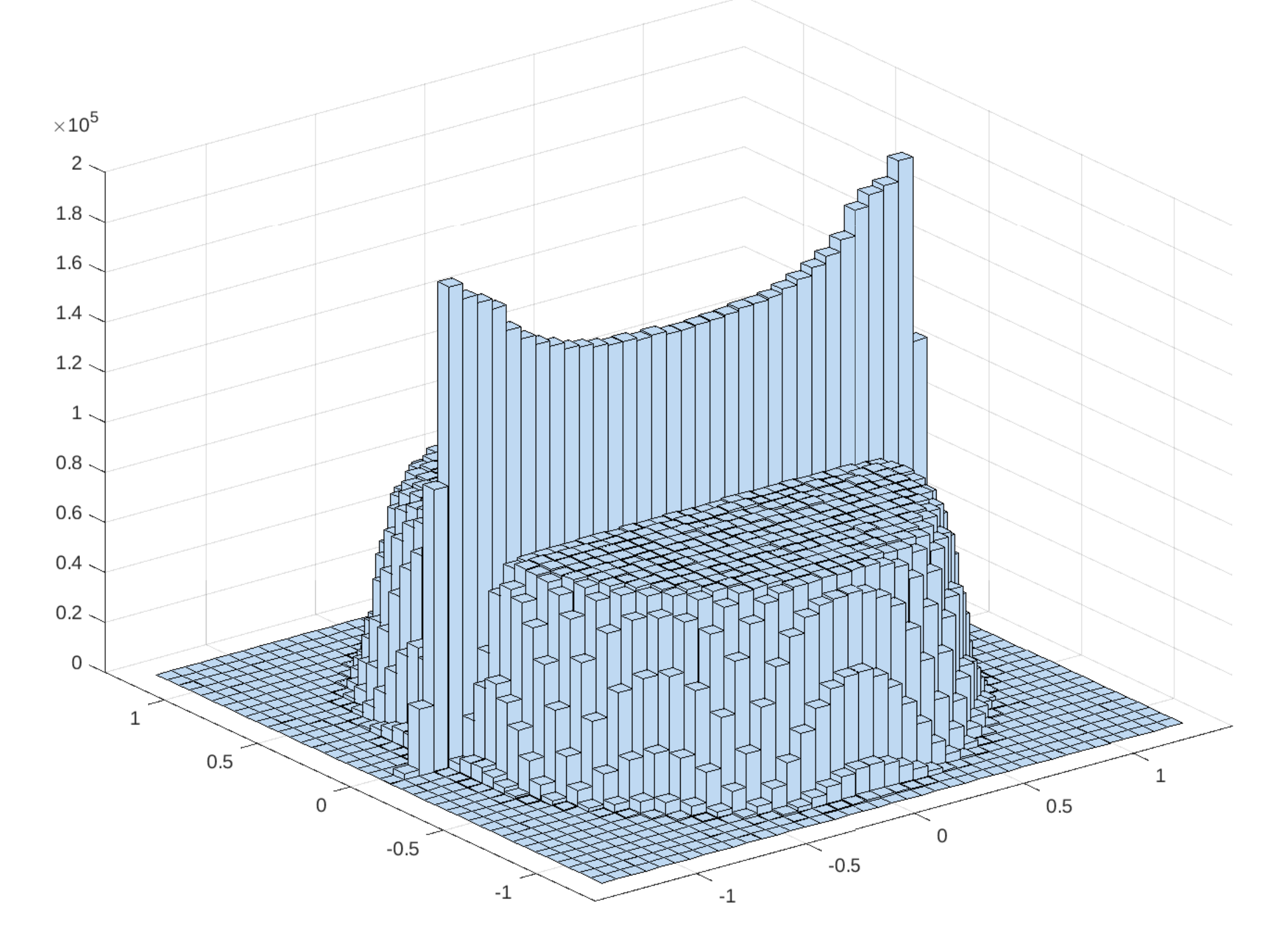}
		
\vspace{0.05cm}
		
\includegraphics[width=0.495\textwidth]{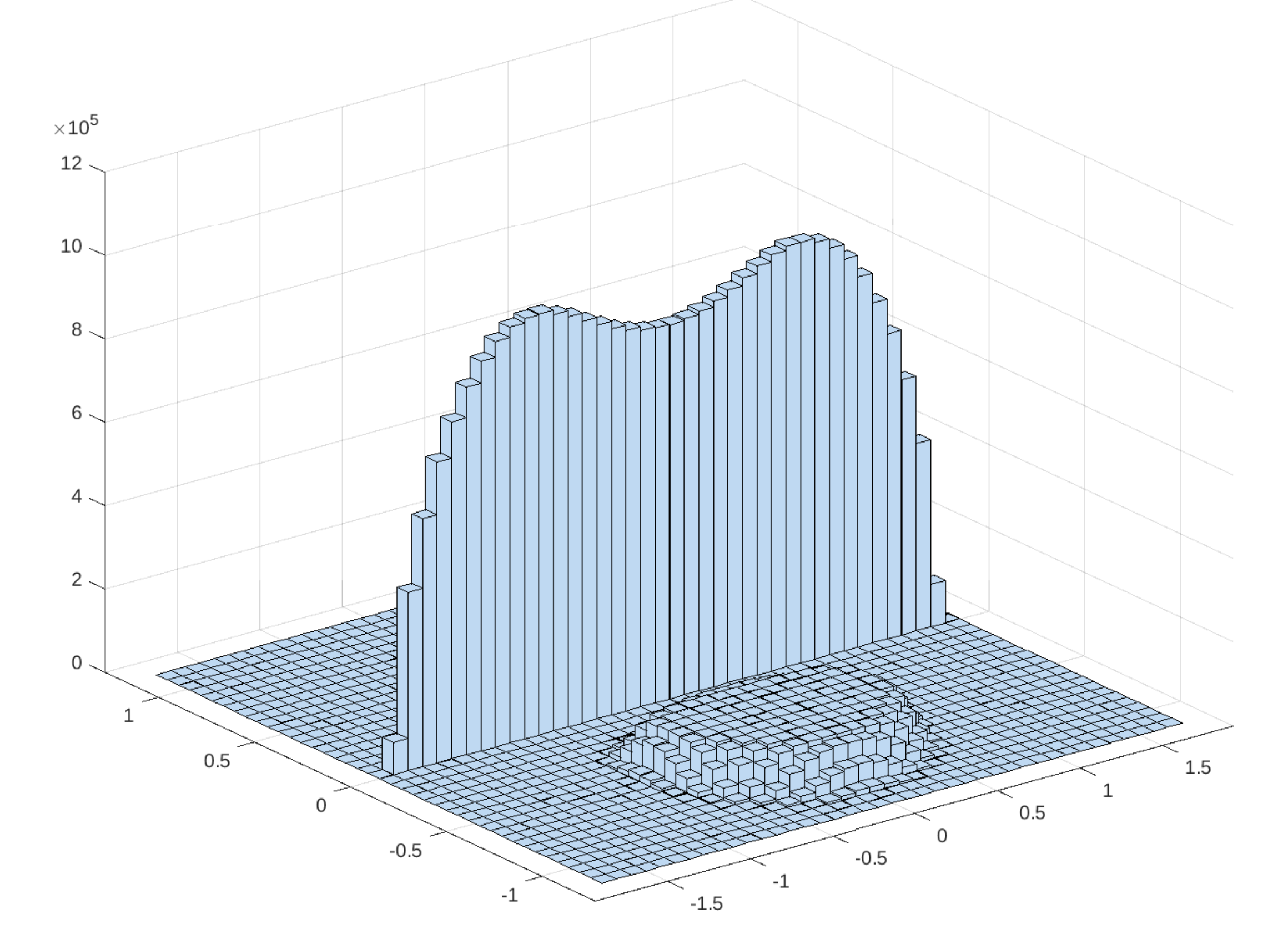}\hfill \includegraphics[width=0.495\textwidth]{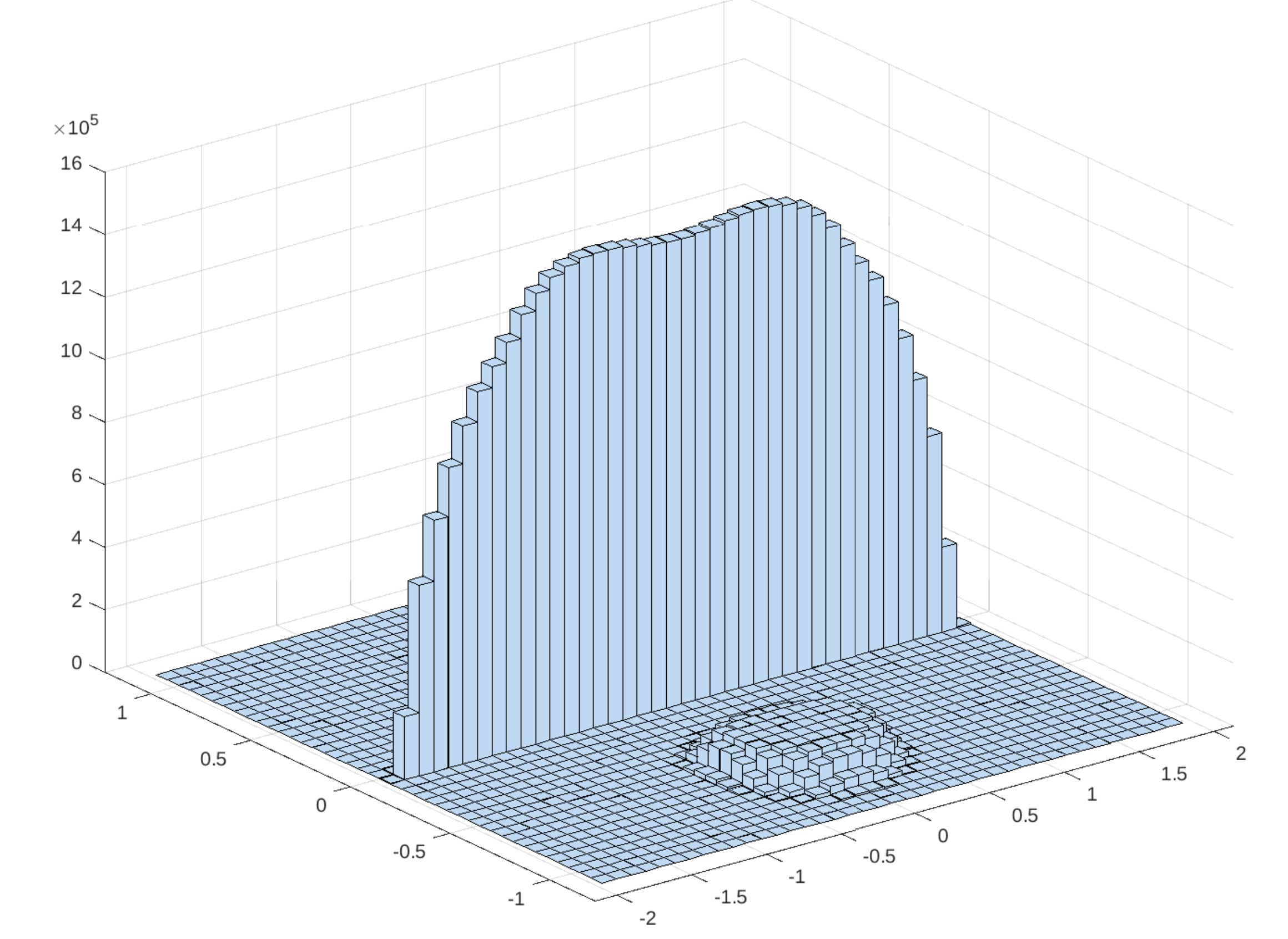}
\end{center}  
\caption{Histograms representing counts of eigenvalues of $\PHI$ for two-dimensional bins in the complex plane showing the eigenvalue distribution obtained by numerical simulations of 500000 samples with $N=128$ for various values of $\lambda$: 1/2, 15/32, 1/4, 1/8 (from upper left to lower right). }\label{fig:histC}
\end{figure}
By Gauss law we can get from the Green's function also the density of complex eigenvalues,
\begin{equation}\label{eq:rhoC}
\rho^{(2)}(w)=\frac{1}{\pi}\frac{\partial}{\partial w^*}G(w)=\frac{m^2}{\pi}
\end{equation}
uniformly on the complex domain and independent of $\lambda$. By taking the product of the area and the density given by \eqref{eq:area} and \eqref{eq:rhoC} respectively, we find that the fraction of complex eigenvalue is given by $1-|1-2\lambda|$, consistent with what we have found on the real axis in Eq.~\eqref{eq:vshaped}.

In Fig.~\ref{fig:histC} we show two dimensional histograms obtained from eigenvalues of many samples of size N=128, for the same lambdas we have used in the scatter plots in Fig.~\ref{fig:scatterB}. In the histograms we counted both real and imaginary eigenvalues, though the two dimensional bins are not well suited for measuring the one-dimensional density and therefore the ratio of the height of the real eigenvalues and the complex eigenvalues is depending on the size of the bins. Even though the theoretical prediction for the degenerated case $\lambda=1/2$ (upper left histogram) is that all eigenvalues are complex, for finite $N$ there are still some real eigenvalues as we have seen in Fig.~\ref{fig:vshaped}. We can see well that inside the bulk the density is very flat. Only at the edges one notices some deviation from the constant density which is a typical finite-$N$ size effect that one can also find in the Ginibre ensemble.

From \eqref{eq:rpm} we find that the complex domain has a distance from the real axis given by
\begin{equation}
m^{-1} \sin(\theta_0/2),
\end{equation}
and it touches the axis when $\lambda$ approaches $1/2$. The latter is the degenerated case which we have mentioned before. Then the complex eigenvalues uniformly fill a disk, like in the (real) Ginibre ensemble.

\subsection{Density of Eigenvalues for $\M{B}=\mathrm{diag}(1,\ldots,1,t,\ldots,t)$}\label{sec:modelt}
In the remaining section of this paper, we want to discuss what happens if we generalize the deterministic metric $\M{B}$ from the last section to the form 
\begin{equation}
\M{B}=\mathrm{diag}(\underbrace{1,\ldots,1}_k,\underbrace{t,\ldots,t}_{N-k}).
\end{equation}
The fraction of ones on the diagonal of $\M{B}$ we still denote by $\lambda$, like in \eqref{eq:lambda}. 
The main motivation for this generalization is that it allows us to study what 
happens when $t$ approaches zero and the metric becomes singular. (The case of singular metric will not be discussed in this paper.) We will concentrate on the case when $t<0$. We shall see that this model exhibits various phases of the density of real eigenvalues and of the complex one. In the two-dimensional space spanned by the parameters $t$ and $\lambda$, phase changes in the system happen on several one-dimensional curves, where we can either describe the critical value of $\lambda$ as a function of $t$, or the critical value of $t$ as a function of $\lambda$. In this section we will variate $t$ while keeping $\lambda$ fixed. Therefore it is more convenient to introduce the critical $t$ as function of $\lambda$. We will denote the different critical values of $t$ by $t_\CC=t_\CC(\lambda)$ for a phase change which is only present in the density of complex eigenvalues, $t_\RR=t_\RR(\lambda)$ when the phase change appears in the density on the real axis, and $t_{\CC\RR}=t_{\CC\RR}(\lambda)$ for a phase change that happens in the density of real and complex eigenvalues at the same time.

Analogously to \eqref{eq:cubic} we get the cubic equation
\begin{equation}\label{eqt:cubic}
v^3+v^2\left(u+\frac{u}{t}\right)+v\left(\frac{u^2}{t}+1\right)+u\left(1-\lambda+\frac{\lambda}{t} \right)=0
\end{equation}
where we have introduced the rescaled variables $u=mw$ and $v=mb$. We can substitute its solution into Eq,~\eqref{eq:Gsol} and find in a similar way as discussed in section \eqref{sec:modelB}, the density on the real axis and the domain of complex eigenvalues.

We are now going to study the change of the complex domain as we keep $\lambda=3/4$ fixed while varrying $t$, starting at the known case $t=-1$ discussed in the previous section. 
\begin{figure}
\begin{center}
\includegraphics[width=0.495\textwidth]{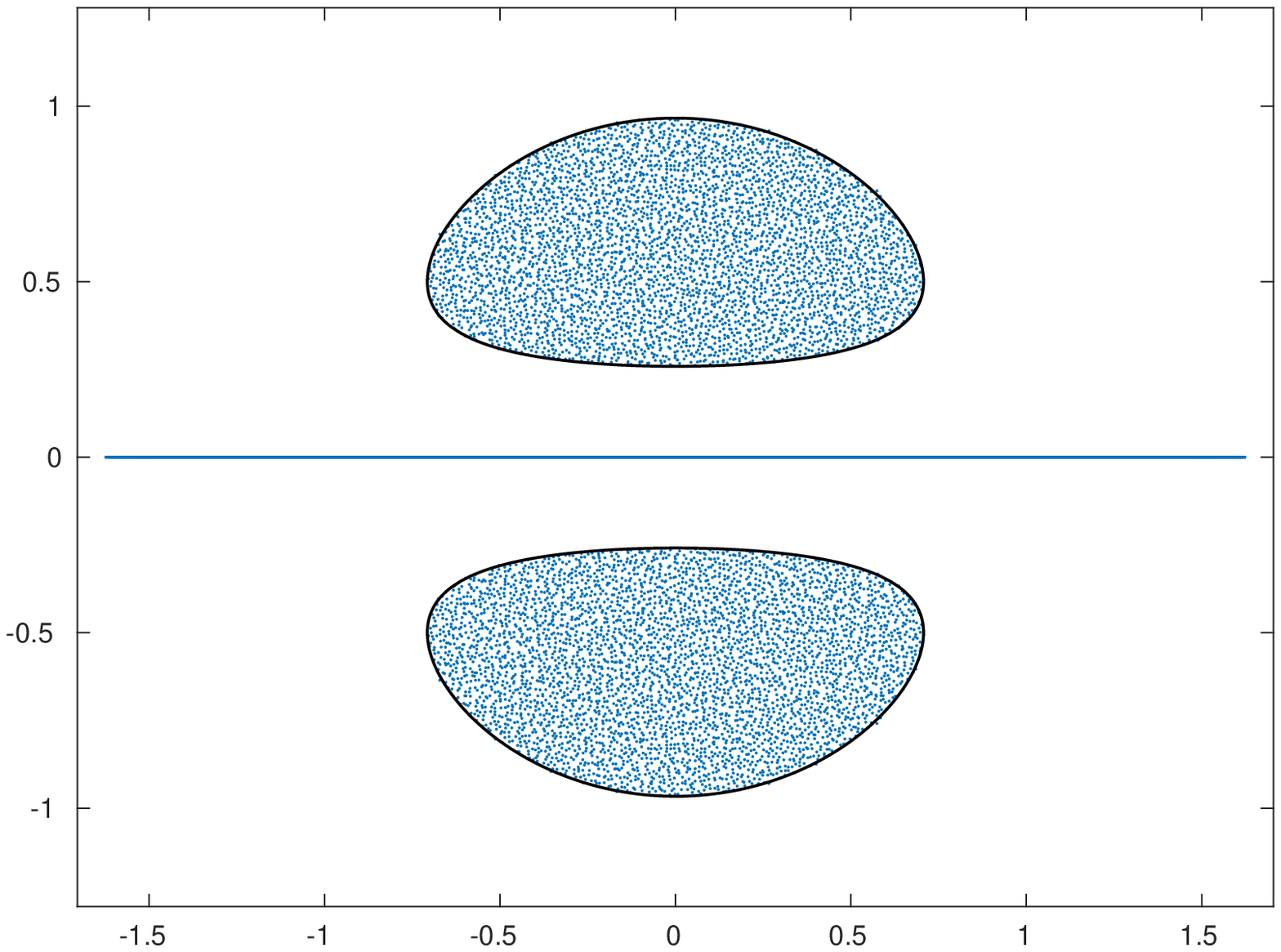}\hfill \includegraphics[width=0.495\textwidth]{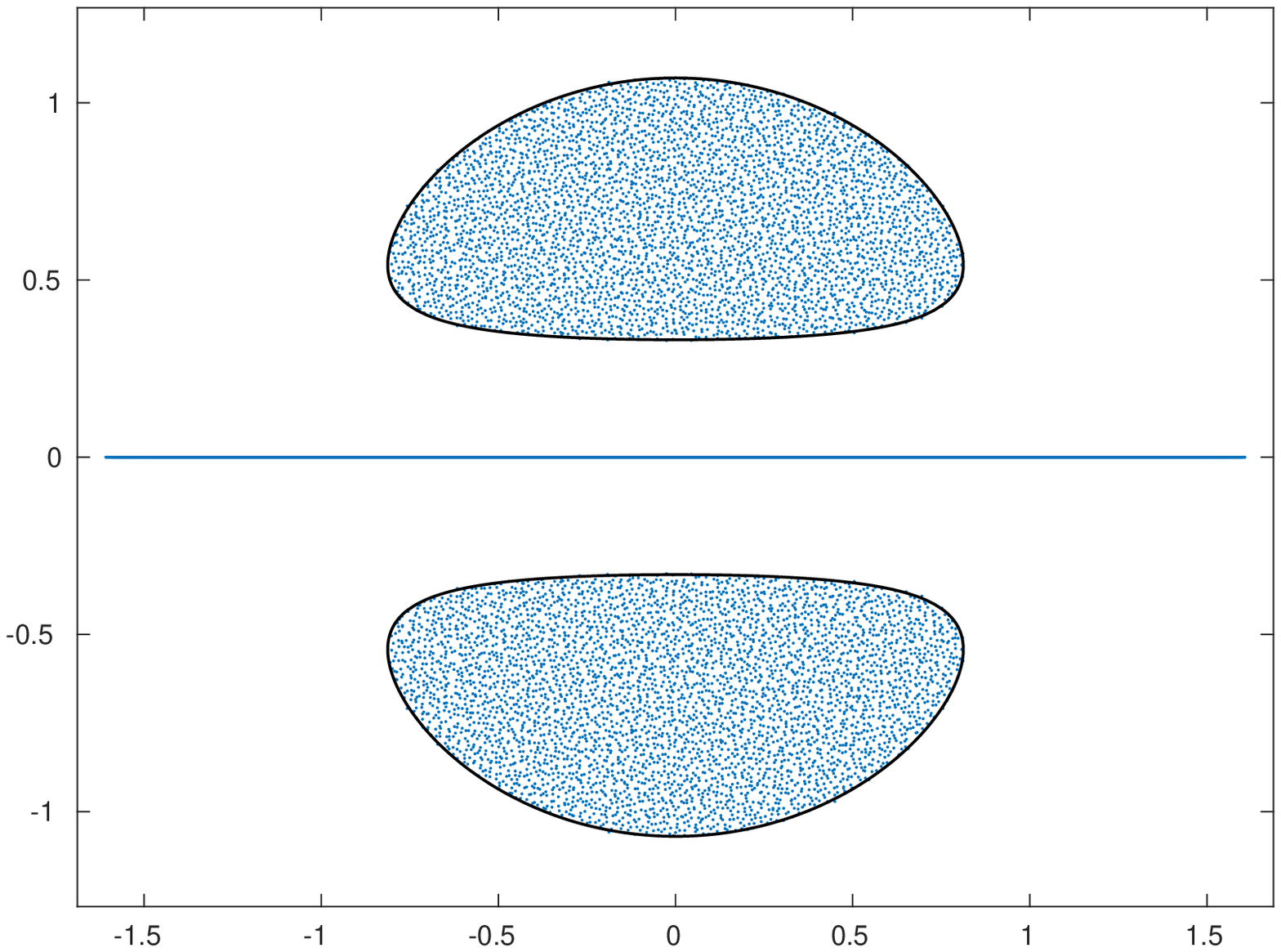}
\end{center}  
\caption{The blue dots represent the eigenvalues of $\PHI=\M{A}\M{B}$ obtained from numerical simulations using a single realization with $N=16384$ and $\lambda=3/4$ for the values of $t=-1$ (left) and $t=-1.2$ (right). The solid black line shows the theoretical boundary when $N\rightarrow\infty$. (The left plot we have put here for comparison. It has already appeared on the left bottom of Fig.~\ref{fig:scatterB} for a different value of $N$.) }\label{fig:tvar}	
\end{figure}
The blue dots in Fig.~\ref{fig:tvar} are again obtained from a scatter plot of a single sample of large size while the black solid lines show the theoretical boundary of the complex domain. We show the effect when we slightly change $t=-1$ corresponding to the model in Section \ref{sec:modelB} to the new model with $t=-1.2$. The boundary of the domain deforms continuously with $t$. The density is still uniform in the complex domain, but it changes with t. Since there is no continuous way for a complex eigenvalue to move onto the real axis, the number of complex eigenvalues does not change, i.e.~the fraction of real eigenvalues is given by $|1-2\lambda|$ as we have found in Eq.~\eqref{eq:vshaped}. 

\begin{figure}
\begin{center}
\includegraphics[width=0.33\textwidth]{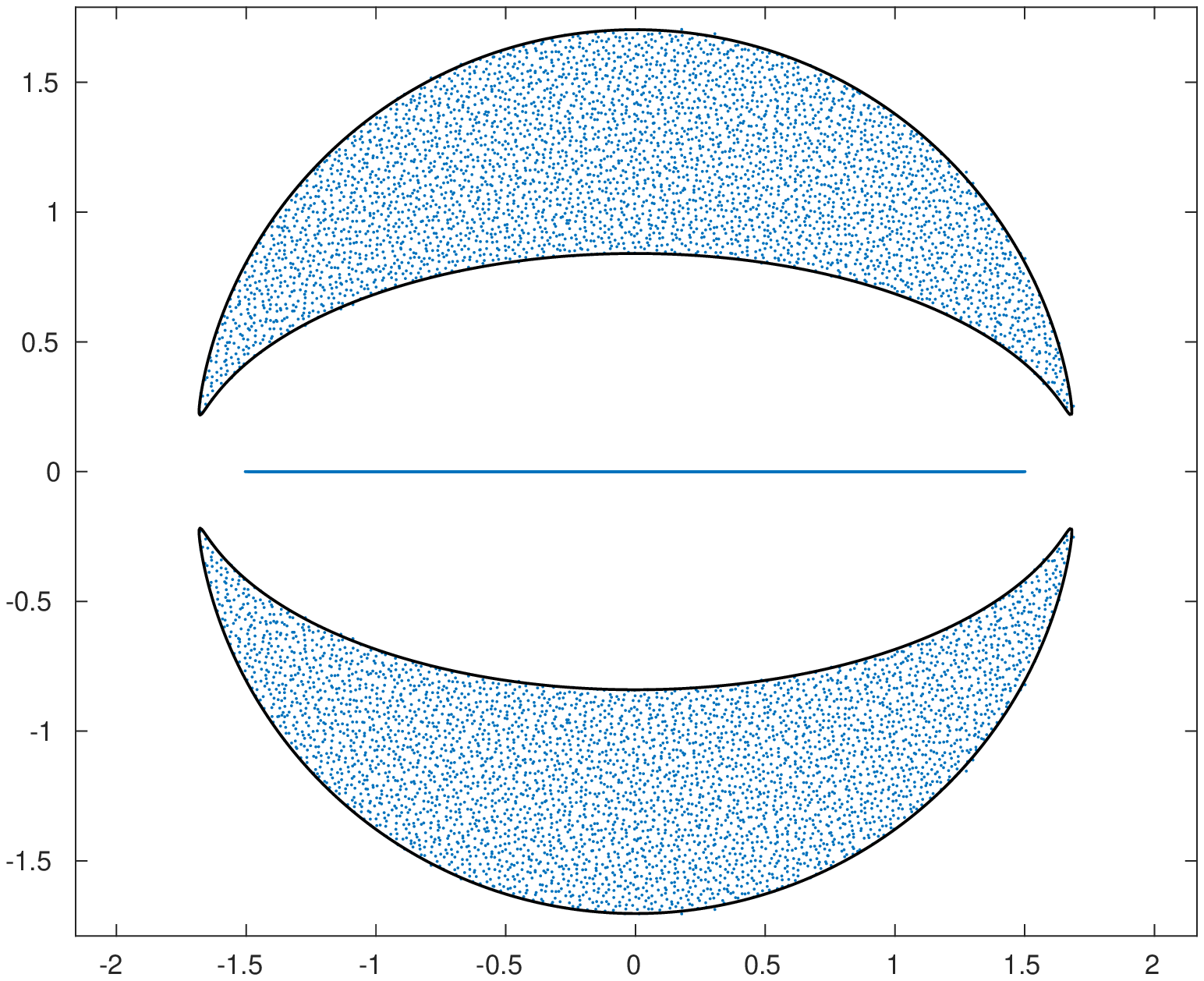}\hfill \includegraphics[width=0.33\textwidth]{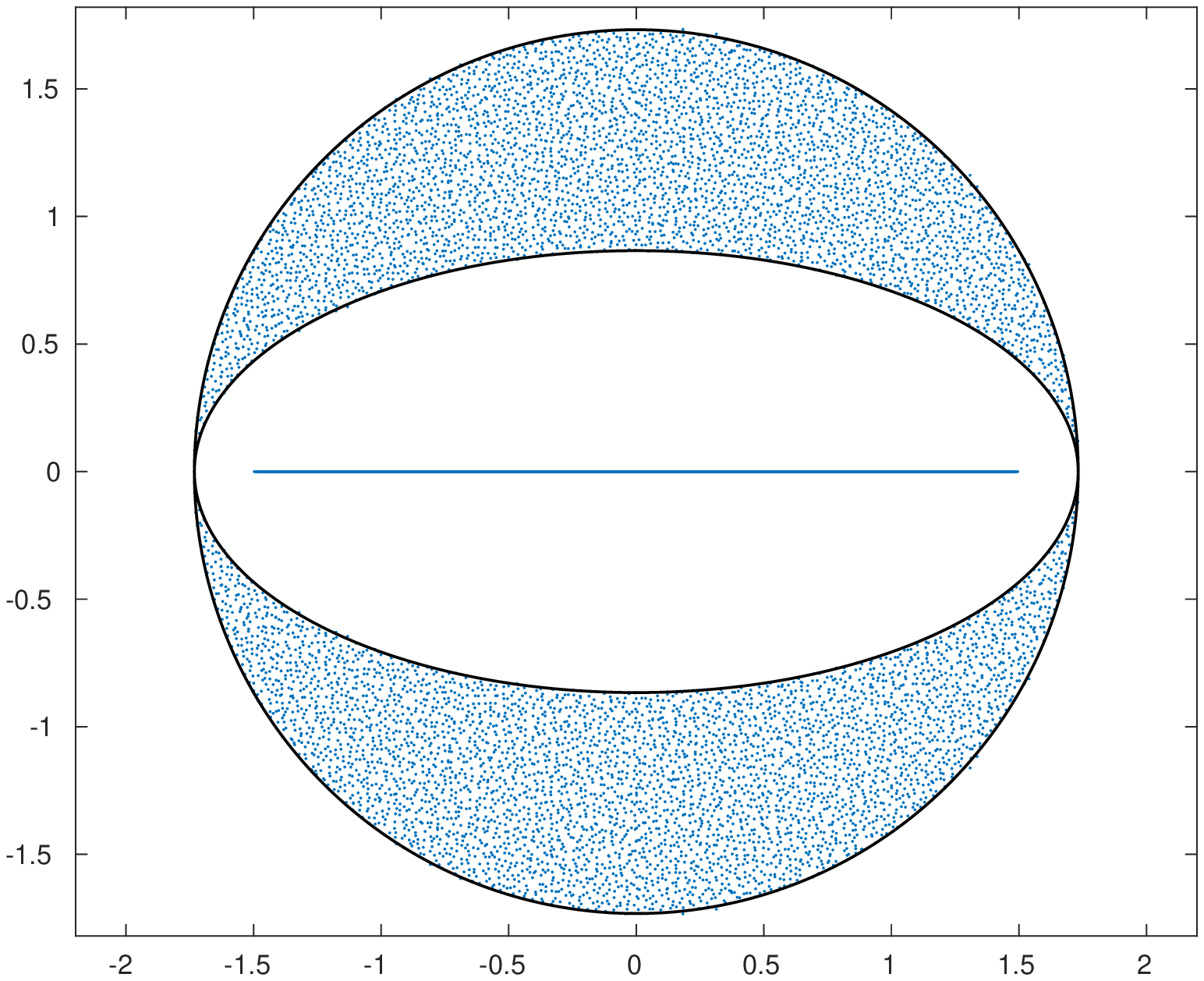}\hfill \includegraphics[width=0.33\textwidth]{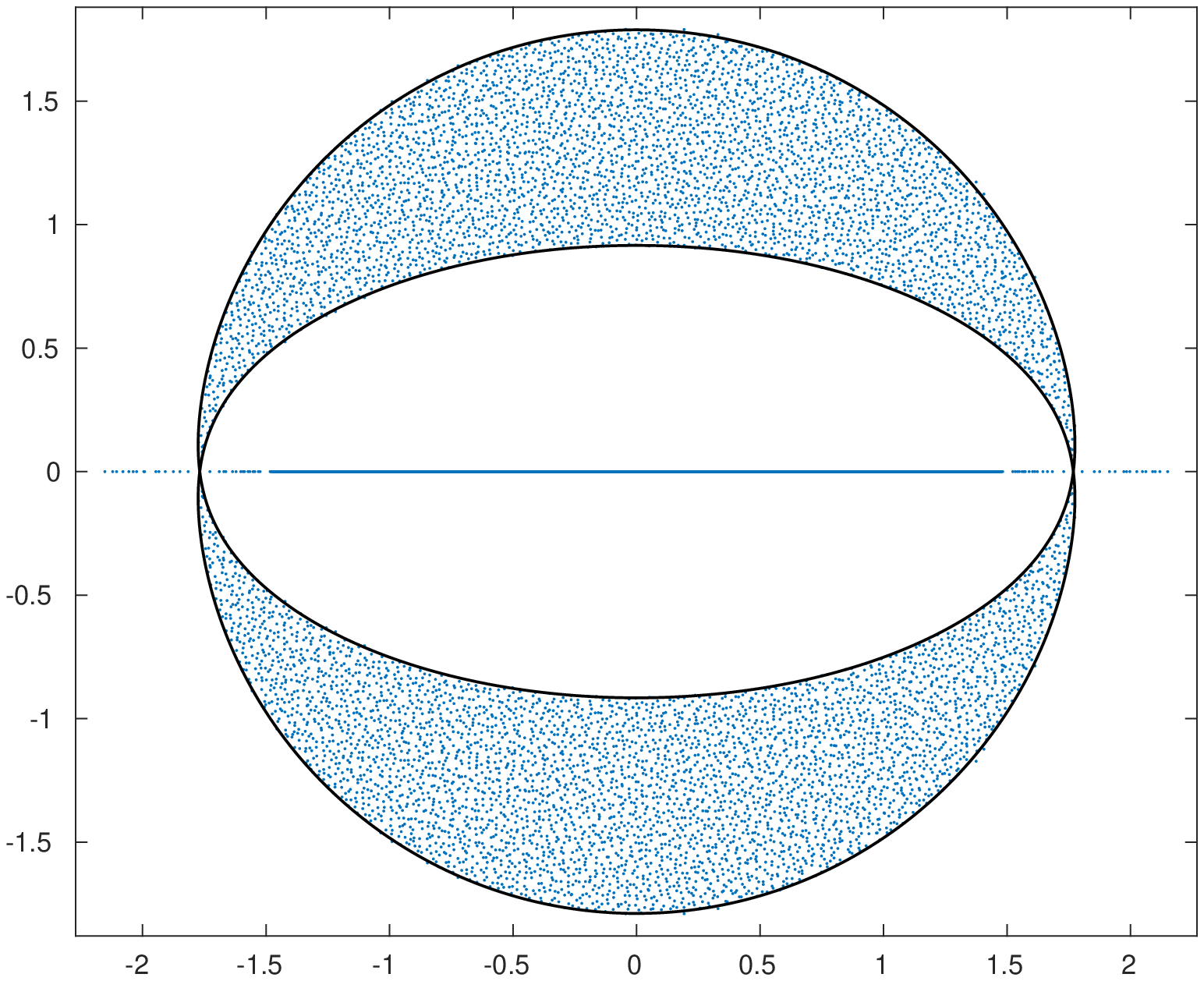}
\end{center}
\caption{The blue dots represent the eigenvalues of $\PHI=\M{A}\M{B}$ obtained from numerical simulations using a single realization with $N=16384$ and $\lambda=3/4$ for various values of $t$: -2.9, -3, -3.2 (from left to right) which represents the pre-critical, critical and post-critical case respectively. The solid black line shows the theoretical boundary when $N\rightarrow\infty$. }\label{fig:tcrit}	
\end{figure}

We are now decreasing $t$ further as presented in  Fig.~\ref{fig:tcrit}, which shows the pre-critical, critical and post-critical cases where $t=-2.9$, $-3$ and $-3.2$, respectively. 
In the first and second case (left and center plot), the fraction of complex eigenvalues is still given by $1-|1-2\lambda|$. In the critical case, when $t=t_{\CC\RR}(\lambda)$, the complex phase changes from a disconnected domain to a connected one as the two crescents touch on the real axis. In the post-critical case (right plot) there is a flow of complex eigenvalues onto the real axis, and the fraction of complex eigenvalues becomes less than $1-|1-2\lambda|$. Since for $t=t_{\CC\RR}(\lambda)$, the crescents touch the real axis outside of the interval of real eigenvalues, on each side a new interval of support of real eigenvalue density emerges in the post-critical phase. 
The fraction of real eigenvalues in the central interval is still given by $|1-2\lambda|$ since at this moment there is no continuous way for the eigenvalues from the outer interval to flow into the central one.

While the total amount of density of real eigenvalues is fixed by $|1-2\lambda|$ when $t\ge t_{\CC\RR}$ (or
the amount of density in the central interval if $t_{\CC\RR} > t > t_{\RR}$), the
endpoints of the interval and the distribution of the density on its support
will still depend on the parameters $t$ and $\lambda$.
Similarly as in Eqs.~\eqref{eq:rhoreal} and \eqref{eq:endpoints} for the model with $t=-1$, one can find
analytic expressions for the density of real eigenvalues and the endpoints of
its support, but they look quite cumbersome. We will state them in an upcoming paper \cite{FR} where we will give more details about the calculation.

\begin{figure}
\begin{center}
\includegraphics[width=\textwidth]{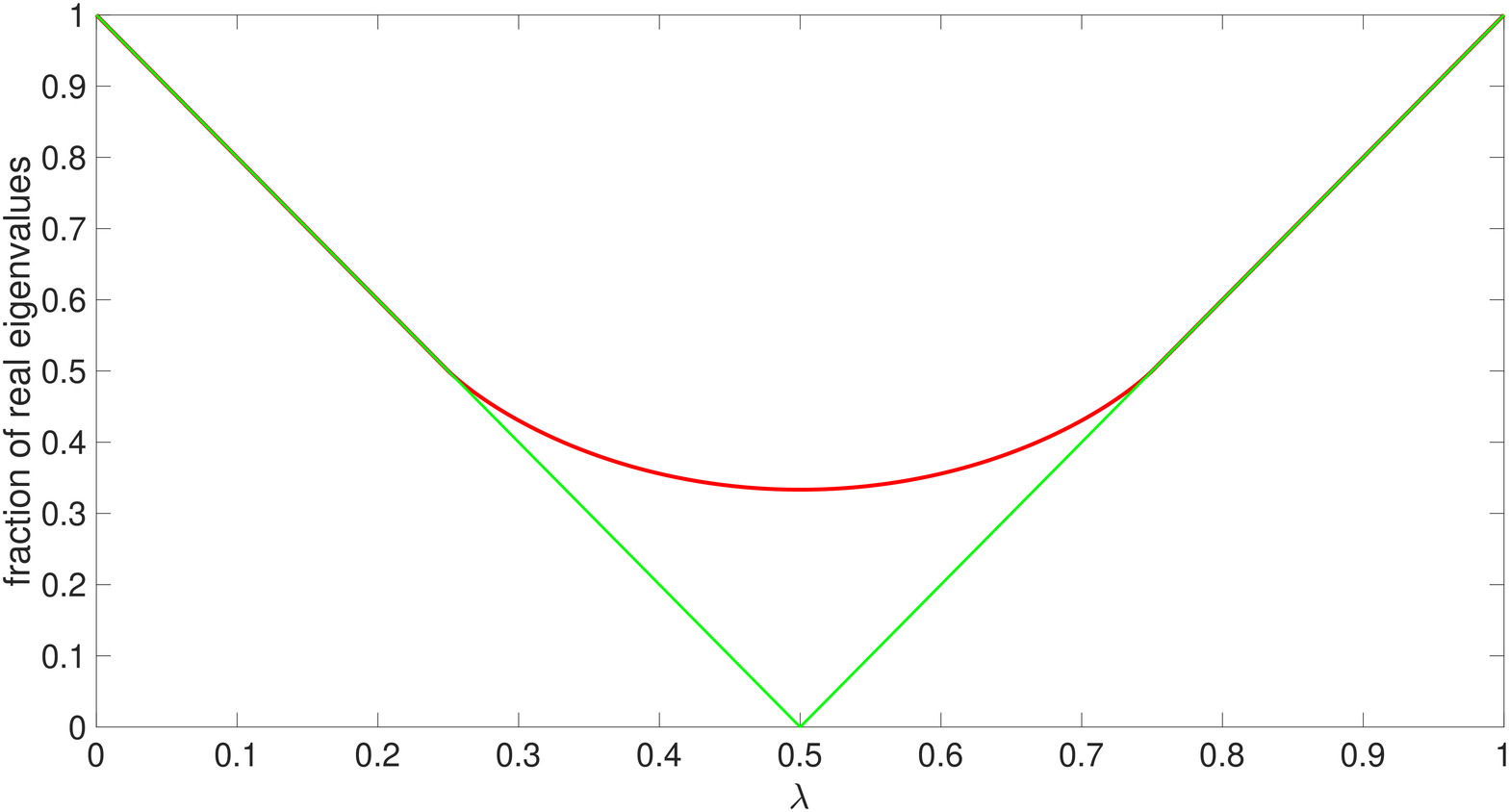}
\end{center} 
\caption{Theoretical large $N$ prediction for the fraction of real eigenvalues as a function of $\lambda$ for $t=-3$ (red) and $t=-1$ (green).}\label{fig:fracR}
\end{figure}
The function that gives the fraction of all real eigenvalues is no longer that simple if $t<t_{\CC\RR}$ and depends on both $\lambda$ and $t$. It is plotted in Fig.~\ref{fig:fracR} as a function of $\lambda$, where we compare the large-$N$ limit for the simpler case $t=-1$ (green) given by \eqref{eq:vshaped} and $t=-3$ (red). 
The critical curves $t_{\CC\RR}(\lambda)$ and $t_{\CC}(\lambda)$ can be also inverted to obtain $\lambda_{\CC\RR}=\lambda_{\CC\RR}(t)$ and $\lambda_{\CC}=\lambda_{\CC}(t)$. For the value $t=-3$ as shown in Fig.~\ref{fig:fracR}, then as long as $\lambda$ is larger than the critical $\lambda_{\CC\RR}=3/4$ (or smaller than $\lambda_\CC=1-\lambda_{\CC\RR}=1/4$) the red curve coincides with the green line. Consistent with the previous discussion, the fraction of real eigenvalues only differs from the simpler case $t=-1$ once we have reached the post-critical phase, where some complex eigenvalues have flown to the real axis. Obviously the fraction is larger than in the simpler case. As was mentioned at the end of the paragraph below \eqref{eq:vshaped}, it follows from \cite{carlson} that the simpler case gives an upper bound for the fraction of complex eigenvalues which again holds for finite $N$ and for each sample.

\begin{figure}
\begin{center}
\includegraphics[width=\textwidth]{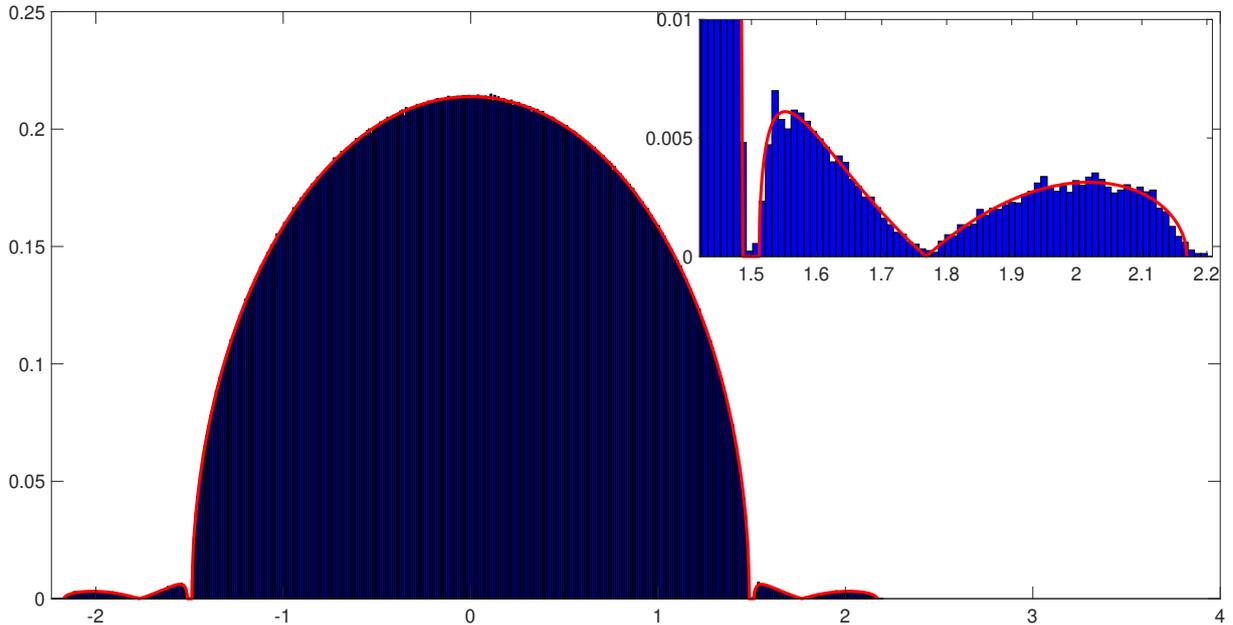}
\end{center}  
\caption{Histogram of real eigenvalues in the post-critical case obtained from 1000 samples with parameters $N=4096$, $\lambda=3/4$, $t=-3.2$. Red solid line shows the theoretical large $N$ result. Inset: Magnification of the right interval.}\label{fig:realt}
\end{figure}
In Fig.~\ref{fig:realt} we show histograms of the density of real eigenvalues in the post-critical case with $t=-3.2$, obtained from 1000 samples. The red solid line shows the theoretical large-$N$ prediction. It shows a small gap between the outer intervals and the central one. Since the gap is very small, it is hard to see it in the
histogram from the numerical simulations, especially since at the edge of the support, the finite-$N$ effects are typically more pronounced than in the bulk. But in the bulk we can see that the red
curve fits perfectly with the histogram. When there are multiple intervals, one can obtain an expression for
the size of the gap by the difference of their endpoints
\cite{FR}.

As can be also seen in Fig.~\ref{fig:realt}, the density on the outer interval also vanishes (on both sides) at a point, without forming a gap in the spectrum. This is the point where the two crescents touch the real line. As this plot shows, in the post-critical case, after the complex domain has started to touch the real line, the density of real eigenvalues in the outer intervals, accounts for all the
complex eigenvalues which have flown to the real axis. When we decrease $t$ even further below $t_{\CC\RR}(\lambda)$, it can happen that the disjoint intervals on the real axis merge again to a single interval. If $1/9<\lambda<8/9$, this merging happens at $t=t_{\RR}(\lambda)$. Otherwise, the endpoints of the intervals will recede towards $\pm\infty$ as $t\rightarrow-\infty$ without merging of the intervals.

\begin{figure}
\begin{center}
\includegraphics[width=\textwidth]{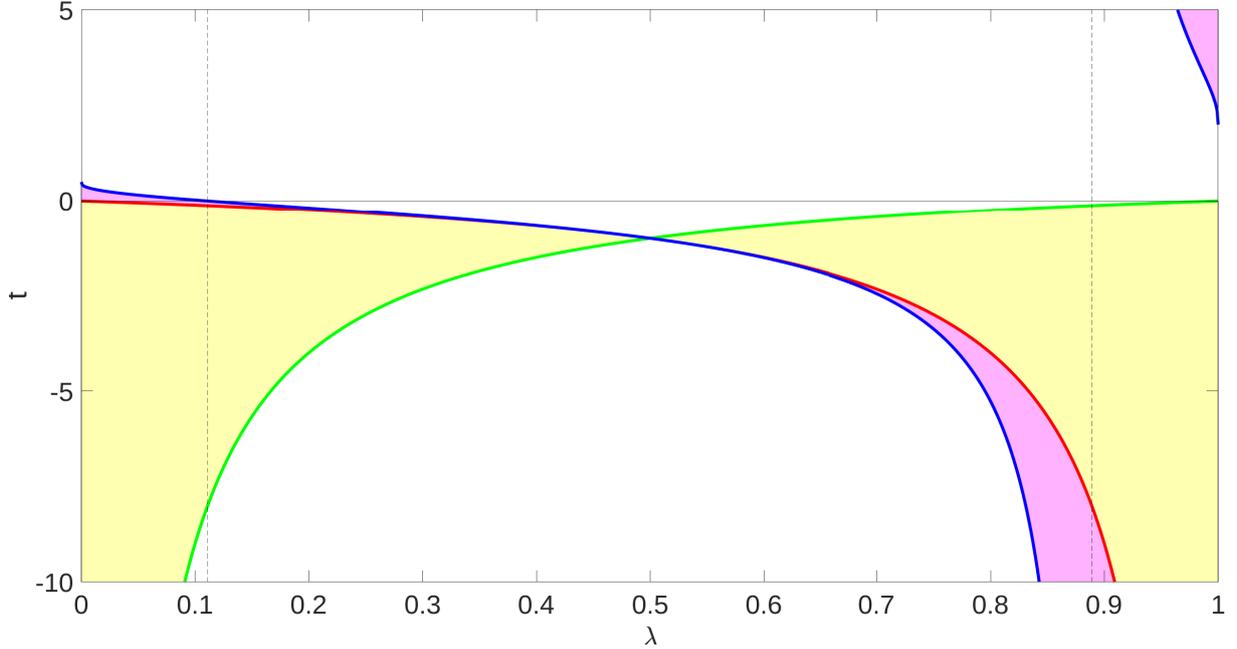}
\end{center}
\caption{Phase diagram of the two-dimensional parameter space of the model where $\lambda$ is on the horizontal and $t$ on the vertical axis. The colored curves represent critical values of the parameters. $t_{\CC}(\lambda)$ (green) given by \eqref{eq:tC} shows the phase transition for complex eigenvalues, $t_{\RR}(\lambda)$ (blue) given by \eqref{eq:tR} for the real eigenvalues when disjoint intervals merge and $t_{\CC\RR}(\lambda)$ (red) given by \eqref{eq:tCR} is a criticality present in real and complex eigenvalues at the same time. For $t<0$ the curve $t_{\RR}(\lambda)$ only exists for $1/9<\lambda<8/9$ which is marked by the vertical dashed lines.
The yellow shaded area represents the phase where the complex domain is disconnected. The magenta shaded area denotes the phase where the line density of real eigenvalues consists of multiple intervals.}\label{fig:phases}
\end{figure}

Finally, in Fig.~\ref{fig:phases}, we summarise the various phases of real and complex eigenvalues of $\PHI$ in a phase diagram of the two-dimensional parameter space. $\lambda$ is along the horizontal axis and $t$ along the vertical one.
Even though in this section we have concentrated on the case where the metric $\M{B}$ is indefinite, i.e.~$t<0$, the discussion below (restricted to the density of real eigenvalues) will still hold in the strictly-quasi-hermitian case when $t$ is positive and there are no complex eigenvalues.  

In all scatter plots in this section we have studied the case for fixed $\lambda=3/4$. Notice that the critical values of $t$ for complex or real eigenvalues depend on $\lambda$. Below we will discuss the various critical curves which can be found in the phase diagram.
 
We will start with the criticality which we have already discussed and shown in Fig.~\ref{fig:tcrit}, when the domain of complex eigenvalues consist of two crescents which touch on the real line. The critical value of $t$ as a function of $\lambda$ is given by
\begin{equation}\label{eq:tCR}
t_{\CC\RR}(\lambda)=\frac{\lambda}{\lambda-1},
\end{equation}
and is plotted in the phase diagram Fig.~\ref{fig:phases} by the red curve. It is both a critical phase for the complex as well as the real eigenvalues since the touching of the crescents happens outside of the interval of real eigenvalues and therefore gives birth to two new disconnected intervals.

There is another criticality in the complex phase given by
\begin{equation}\label{eq:tC}
t_{\CC}(\lambda)=\frac{\lambda-1}{\lambda}=\frac{1}{t_{\CC\RR}(\lambda)},
\end{equation}
represented by the green curve in the phase diagram. In contrast to the previous case it is not a criticality for the real eigenvalues.
In this phase the domain of complex eigenvalues is convex and the touching point with the real axis happens at the origin. Since at this moment, there is some density of real eigenvalues in the neighborhood of the origin, the complex eigenvalues can flow in a continuous way to the existing interval of real eigenvalues.

Finally there is the criticality which only concerns the real eigenvalue density, given by    
\begin{equation}\label{eq:tR}
t_{\RR}(\lambda)=\frac{1}{3(8-9\lambda)}\left( 6(2-3\lambda)-\frac{|\xi_\lambda+\sqrt{\Delta_\lambda}|^{1/3}+|\xi_\lambda-\sqrt{\Delta_\lambda}|^{1/3}}{2^{1/3}} \right),
\end{equation}
where 
\begin{equation}
\Delta_\lambda=531441(8-9\lambda)^2(1-\lambda)^2\lambda^2\ge 0,
\end{equation}
and 
\begin{equation}
\xi_\lambda=-729\lambda(\lambda-1)(7\lambda-8).
\end{equation}
In the phase diagram, $t_{\RR}(\lambda)$ is plotted in blue. This critical curve is associated with the merging of three disjoint intervals of real eigenvalues into a single one. If we restrict $t$ to be negative, it only exists when $1/9<\lambda<8/9$ which is marked by the vertical dashed lines in Fig.~\ref{fig:phases}. The right vertical line is an asymptote of the blue curve.

If for fixed $\lambda$, $t$ lies between $t_\CC(\lambda)$ and $t_{\CC\RR}(\lambda)$, i.e.~in the yellow shaded area in the phase diagram, the domain of complex eigenvalues is disconnected. Else, i.e.~when $t$ lies in the white or magenta shaded areas, the complex domain always touches the real axis. When $t$ lies between  $t_{\RR}(\lambda)$ and $t_{\CC\RR}(\lambda)$, i.e.~in the magenta shaded area, the density of real eigenvalues consist of three disjoint intervals, while else it is supported on a single interval. 

The simple law where the fraction of real eigenvalues as a function of $\lambda$ is given by $|1-2\lambda|$ is only valid in the yellow shaded area of the phase diagram, while it still holds for the fraction of real eigenvalues in the center interval in the magenta shaded phase.  

All the critical lines, in red, blue, and green, intersect at $\lambda=1/2$ and $t=-1$. Because of this, we have seen only one critical case in the simpler model in Section \ref{sec:modelB} where $t=-1$. There it corresponds to the degenerated case where the upper half disk touches the lower one.

\section{Discussion} 
In this work we reviewed our recent results on the spectral properties of large random matrices which model systems whose hamiltonians are either quasi-hermitian or pseudo-hermitian. Large quasi-hermitian or pseudo-hermitian matrices can be thought of truncations of the corresponding operators to finite dimensional vector spaces. Thus, in the limit of large matrix size, one may gain useful knowledge on the corresponding operators themselves. In the usual spirit of Random Matrix Theory \cite{Mehta}, randomness of these matrices can model complicated or chaotic systems, or systems which are disordered to begin with. As explained in detail in \cite{FRDecember, MK}, in the quasi-hermitian case, the powerful methods of free probability theory can be brought to bear to deriving analytically an explicit  expression for the average density of vibrational modes of a highly connected large disordered mechanical system. An important physical feature of such systems is a universal behaviour of the density of modes at low frequencies, namely, that the density of low frequency phonons tends to a non-vanishing constant.  This is also the typical behaviour of the phonon spectrum of a regular (periodic) one-dimensional chain of atoms. Thus, in more formal words, one may say that such highly connected  mechanical systems universally have spectral dimension one. 
The bulk of this review was dedicated to the pseudo-hermitian case \cite{FRB}. We have applied the diagrammatic method to obtain an explicit analytical expression for the average density of eigenvalues of such matrices both on the real axis and in the complex plane and discovered a rich phase structure of this object. 
We have also carried meticulous numerical analysis of these matrices. The numerical results agree very well with our analytical predictions. 
The results presented here are just the tip of the iceberg. Quasi- and pseudo-hermitian random matrices are clearly a promising new direction in the vast ocean of Random Matrix Theory. 

\ack
This research was supported by the Israel Science Foundation (ISF) under grant No. 2040/17.  Computations presented in this work were performed on the Hive computer cluster at the University of Haifa, which is partly funded by ISF grant 2155/15. Finally, we thank T.~Can for turning our attention to \cite{carlson}.

\end{document}